\documentclass[aps,prd,twocolumn,superscriptaddress,amssymb,eqsecnum,showpacs,showkeyes,secnumarabic,graphics,floatfix,nofootinbib,tightenlines,longbibliography]{revtex4-1}
\usepackage{graphicx}
\usepackage{bm}
\usepackage{float}
\usepackage{subfigure}
\usepackage{amsmath}
\usepackage{amsfonts}
\usepackage{epsfig}
\usepackage{mathrsfs}
\usepackage[utf8]{inputenc}
\usepackage[breaklinks=true,colorlinks=true,
linkcolor=blue,urlcolor=blue,citecolor=cyan, 
bookmarks=true,bookmarksopenlevel=2]{hyperref}

\restylefloat{figure}

\def\bed{\begin{description}}
\def\eed{\end{description}}

\def\ba{\begin{array}}
\def\ea{\end{array}}

\def\half{\frac{1}{2}\,}

\begin{document}
\newcommand{\newc}{\newcommand}
\newc{\be}{\begin{equation}}
\newc{\ee}{\end{equation}}
\newc{\bear}{\begin{eqnarray}}
\newc{\eear}{\end{eqnarray}}
\newcommand{\mincir}{\raise
-3.truept\hbox{\rlap{\hbox{$\sim$}}\raise4.truept\hbox{$<$}\ }}
\newcommand{\magcir}{\raise
-3.truept\hbox{\rlap{\hbox{$\sim$}}\raise4.truept\hbox{$>$}\ }}

\title{Constraints on scalar coupling to electromagnetism}

\author{Ioannis Antoniou}\email{ianton@cc.uoi.gr}
\affiliation{Department of Physics, University of Ioannina, GR-45110, Ioannina, Greece}

\date{\today}

\begin{abstract}

We review a possible non-minimal coupling (dilatonic) of a scalar field (axion like particle) to electromagnetism, through experimental and observational constraints. Such a coupling is motivated from recent quasar spectrum observations that indicate a possible spatial and/or temporal variation of the fine-structure constant. We consider a dilatonic coupling of the form $B_F(\phi)=1+g\phi$. The strongest bound on the coupling parameter $g$ is derived from weak
equivalence principle tests, which impose $g<1.6 \times 10^{-17} GeV^{-1}$. This constraint is strong enough to rule out this class of models as a cause for an observable cosmological variation of the fine structure constant unless a chameleon mechanism is implemented. Also, we argue that a similar coupling occurs in chameleon cosmology, another candidate dark mater particle and we estimate the cosmological consequences by both effects. It should be clarified that this class of models is not necessarily ruled out in the presence of a chameleon mechanism which can freeze the dynamics of the scalar field in high density laboratory regions.
\end{abstract}
\maketitle

\section{ Introduction}

There are recent observational indications that the fine structure
constant may be varying spatially and/or temporally
\cite{Webb:2010hc,
Chiba:2011bz,Murphy:2003hw,Mota:2003tm,Reinhold:2006zn,Fenner:2005sr,Langacker:2001td}
on cosmological scales. Such a variation could be due to a scalar
field non-minimally coupled to electromagnetism. This field
could also play the role of quintessence inducing the observed
accelerating expansion of the universe
\cite{Marra:2005yt,Nesseris:2004uj,Anchordoqui:2003ij,Tsujikawa:2013fta}.
The possible spatial variation of the fine structure constant
would require a corresponding spatial variation of the scalar
field which could be supported by non-trivial topological
properties of the field configuration
\cite{Amendola:2011qp,Platis:2014sna,Perivolaropoulos:2014lua,Mariano:2012ia}.
The variation of the fine structure constant is given by the relation 
$\frac{\Delta\alpha}{\alpha}=\frac{\alpha-\alpha_0}{\alpha_0}$, where $\alpha_0$ 
is the present value and it is of order $O(10^{-5})$ \cite{King:2012id} (for a spatial variation).

We focus on cases where scalar particles or chameleons are subject
to coupling with the electromagnetic tensor \cite{Raffelt:1987im}.
We consider an Lagrangian interaction term of the form: 
\be L_{coupling}=-\frac{1}{4} B_F(\phi) F_{\mu \nu} F^{\mu\nu}\label{glagra}\ee  where: \be B_F(\phi)= 1 + g\phi\label{bf}\ee 
is the gauge kinetic function, $\phi$ is a scalar
field such as axion-like particle (ALP), chameleon, quintessence, etc and $g$ is
the coupling constant, which must be constrained. 

Axions are particles, whose existence helps to solve the strong CP problem. Also, they are dark matter candidate particles  because they interact mostly gravitational and can induce the required dark matter density of the universe. Their mass and their coupling to electromagnetism is constrained by laboratory, cosmological and astrophysical bounds \cite{Duffy:2009ig}. 

ALPs are dark matter candidates \cite{Arias:2012az,Cadamuro:2012rm} (section 2). For consistency with the observed accelerating expansion rate, the required magnitude of the coupling $g$ of the scalar field is described for example in Ref. \cite{Calabrese:2013lga,Calabrese:2011nf}. It is therefore interesting to inquire if such values of the coupling are consistent with local experiments and astrophysical observations. This is the goal of the present analysis.

In the next section we present experimental bounds from the photon-ALP coupling and a brief discussion about these experiments while in section $III$ we present
corresponding bounds for chameleon scalar fields. In section $IV$ we discuss astrophysical and cosmological constraints on $g$ and in
section $V$, we conclude and summarize.

\section{  Constraints from photon-axion like scalar coupling}

We focus on the class of experiments designed to constrain or
detect the interaction between scalar ALPs and
photons. Generally, a positive signal can determine mass, parity
and coupling of the hypothetical scalar particle. This coupling,
if ALPs are scalar, is described by the Lagrangian term \cite{Chou:2007zzc}:
\be{L_{scalar}}=-\frac{g}{4}\phi F_{\mu\nu} F^{\mu \nu}\label{lagra}\ee 
where: 
\be F_{\mu \nu} F^{\mu\nu}=2 ({\textbf{B}}^2-{\textbf{E}^2})\ee 
If we compare the relations (\ref{glagra}), (\ref{bf}) and (\ref{lagra}), it is easy to show that: 
\be{L_{coupling}}={L_{scalar}}-\frac{1}{4}F_{\mu\nu} F^{\mu \nu} \ee 
 If ALPs are pseudoscalar, the corresponding Lagrangian term is:
\be{L_{pseudoscalar}}=-\frac{g}{4}\phi F_{\mu \nu} \widetilde{F}^{\mu \nu}\ee
where: 
\be  F_{\mu \nu} \widetilde{F}^{\mu \nu}= -2{\textbf{E}}\cdot{\textbf{B}}\ee 
The quantity $\widetilde{F}^{\mu\nu}$ is the dual electromagnetic tensor which violates parity and
time reversal invariance. It conserves charge conjugation
invariance, so it violates CP symmetry. In both cases, the
expression for the coupling between ALPs and
photons is given as \cite{Masso:1995tw}: 
\be g \equiv\frac{1}{M}\approx\frac{\alpha}{2\pi}\frac{m}{f_\alpha}\ee 
where $\alpha\simeq 1/137$ is the fine structure constant, $m$ the mass
of the scalar or pseudoscalar particle and $f_\alpha$ the symmetry
breaking scale (or decay constant). As the decay constant
increases, the coupling $g$ decreases. However, it  can not be
greater than $f_\alpha \sim 10^{-16} GeV$ \cite{Hertzberg:2008wr},
because this would lead to closed universe.

ALPs can have odd (pseudoscalar) or even
(scalar particles) parity and can couple to two photons. There are
four classes of experiments attempting to detect such particles.
The first is based on the so-called haloscope
\cite{Asztalos:2003px}. In this experiment, ALPs from galactic halo, converts to photons in a
cavity with a powerful magnetic field. The second
category comes from the so-called helioscope
\cite{Zioutas:2004hi}, which corresponds to weakly interacting slim particles
(WISPs) emitted by the Sun. The third class involves searching for
ALPs which couple to photos and induce in a laser beam,
which propagates in a magnetic field, optical dichroism and
birefringence \cite{Zavattini:2005tm}. The fourth class includes photon
regeneration experiments \cite{Redondo:2010dp}, such as GammeV \cite{Chou:2007zzc}, BFRT \cite{Ruoso:1992nx},
OSCAR \cite{Pugnat:2007nu} and others described bellow. A possible
signal currently exists from the third class of experiments
(PVLAS). For a brief but not complete review, see Ref.
\cite{Januschek:2014tua}.

Most of these experiments are based on fundamental optical
properties of the materials affecting their interaction with polarized
light, such as \cite{Raffelt:1996wa,born1999principles}:
\begin{itemize}
    \item \textbf{optical rotation (activity)}:  is the turning of the plane of
    linearly polarized light about the direction of motion as the light
    travels through materials. It is due to a selective attenuation
    of one polarization component \cite{Zavattini:2005tm}.
    \item \textbf{birefringence}: is the optical property of a material having a
    refractive index that depends on the polarization and direction of light propagation
    \cite{GarciadeAndrade:2001ii}. The birefringence is often quantified as the
    maximum difference between refractive indices exhibited by the material.
    \item \textbf{dichroism}: there are two related but distinct meanings \cite{Villalba-Chavez:2013goa}.
    Dichroism is the phenomenon where  light rays, having different
    polarizations, are absorbed by different amounts, or where a
    visible light can be split up into distinct beams of different
    wavelengths \cite{Rubbia:2008us,Antoniadis:2006wp}.
    \item \textbf{ellipticity}: is the phenomenon where  the polarization of
    electromagnetic radiation, such that the tip of the electric field
    vector, describes an ellipse in any fixed plane intersecting the direction
    of propagation \cite{Dinu:2014tsa}. It is due to selective retardation of
    one polarization component. In that case the direction of the rotation, and
     thus the specified polarization, may be either clockwise or counter
     clockwise.
\end{itemize}

 \subsection{ PVLAS experiment}

The PVLAS experiment takes place at the INFN Legnaro National
Laboratory, near Padua in Italy. In $2006$ they reported
a positive signal for a zero-spin particle \cite{Zavattini:2005tm}.
This experiment is based on the fact that vacuum, in the presence of the scalar field, becomes
birefringent and dichroic \cite{Antoniadis:2007sp} when applying
an external magnetic field \cite{Ahlers:2006iz}. So, when a linear
polarized beam propagates in a Fabry-Perot cavity with strong magnetic
field, the plane of polarization is rotated by an angle $\alpha$.

The  polarized laser beam has wavelength $\lambda=1064 nm$ or
$\lambda=532 nm$ and enters in a high transverse magnetic field of order $5T$, in a cavity. 
It passed $44000$ times through a $1 m$ long magnet. The
components of the laser polarization had a slight weakening. This
effect is observed at varying levels if the polarization is
transverse or parallel to the external magnetic field. The rotation angle was found to be: 
\be \alpha=(3.9 \pm 0.5)\times 10^{-12} rad/pass\ee  
The signal was associated to a neutral, light boson produced by a two-photon
vertex. The amplitude of the dichroism, which depends on the
coupling constant $g$, was estimated as \cite{Chou:2007zzc}: 
\be g\sim 2.5\times 10^{-6} GeV^{-1}\ee  
The mass of the particle was estimated as $m_\phi\sim 1.2 meV$ but its parity was undetermined,
although the sign of the phase shift hints towards even parity (scalar).

This signal could also be explained by assuming the existence of
millicharged particles \cite{Melchiorri:2007sq}. They are light
particles with electric charge $q\ll e$, where $e$ is the
elementary (electron/proton) charge and appear in field theories, but
they aren't part of the Standard Model \cite{Davidson:2000hf}.
The PVLAS experiment was repeated without detection of any signal
\cite{Zavattini:2007ee}. Thus, its results are currently under question.

 \subsection{ GammeV experiment}

The GammeV experiment \cite{Chou:2007zzc} takes place at
Fermilab and consists of two similar experiments. They are 'light shining through a wall'
experiments based on the Primakoff effect, where two photons
with high energy interact and produce ALP. One
photon is real from the laser field and the other one is virtual
from an external magnetic field.

The Primakoff effect is the production of bosons, when high energy
photons interact with an atomic nucleus. Also, include the rotation of the
plane of polarization, when a linearly polarized beam passes
through a magnetic field. The beam has many directions of polarization.
The Primakoff effect reduces  the  parallel component  
of polarised  light  to the  magnetic field and leaves the perpendicular 
component  to the  magnetic  field unchanged. This
phenomenon can occur in a reverse manner (a particle can decay
into two photons).

The GammeV experiment is a gamma ($\gamma$) to milli-eV ALP search. The mass of this particle is expected to be of
order $meV$. A scalar particle couples to photons with a polarization orthogonal to the
magnetic field \cite{Redondo:2010dp}, \cite{Brax:2012ie}. The photon beam is blocked by the wall, but the
ALPs hardly interact with the wall and passes
through the wall. The particles convert again to photons in the
magnetic field and the regenerated photons are counted with an
appropriate detector. The primary and the regenerated
photons have the same properties. The photon
regeneration experiment is based on different effects of light,
compared to the optical rotation experiment. In the first, the
appearance of light beyond the wall is detected, while in the
second, perturbations of the initial beam are detected.

The photon to scalar particle conversion probability (and the
reverse process), is given by the relation: 
\be P_{\gamma\leftrightarrow s}= \frac{1}{4u}(g B L\sin\theta)^2(\frac{2}{q L}\sin\frac{qL}{2})^2\label{probability}\ee 
where the transverse magnetic field $B$ has length $L$, and $\theta$ is the angle between the laser
polarization and the magnetic field. It is clear that, the
direction of polarization must be perpendicular to magnetic field
for optimum conversion. For pseudoscalar particles it must be
parallel to magnetic field, because the probability contains the
term $\cos\theta$ instead of $\sin\theta$. Here, $g$ is the
coupling constant, $u$ the velocity of the scalar particle and $q$
the momentum transfer. The probability becomes maximum when $q
\cdot L\rightarrow0$, ie when the particle has very little mass
compared to its energy ($m \ll \omega$). In order to increase the convention probability, we must use strong,
long range magnetic fields. The momentum transfer is proportional to the square mass of the particle: 
\be q= \mid \omega-\sqrt{\omega^2-m^2}\mid \simeq\frac{1}{2}\frac{m^2}{\omega}\ee.

We can split the above probability (\ref{probability}) in two phases. The first one is the probability in the production region: 
\be P_{\gamma \rightarrow s}= \frac{(2g B \omega sin\theta)^2}{m_\alpha^4}(\sin\frac{L_1m_\alpha^2}{4\omega})^2\label{probability1}\ee 
(where the photons convert to scalar particles \cite{Jaeckel:2010ni}), which increases with the number of passes through the wall.
The second, is the probability in the regeneration region: 
\be P_{s\rightarrow \gamma }= \frac{(2g B \omega sin\theta)^2}{m_\alpha^4}(\sin\frac{L_2 m_\alpha^2}{4\omega})^2\label{probability2}\ee 
(where the scalars reconvert to photons), which increases by using a resonant cavity in the regeneration region.
The expected counting rate of photons in the detector is of the form: 
\be \frac{dN_\gamma}{dt}=\frac{P}{\omega}\eta (P_{\gamma\leftrightarrow s})^2\label{rate}\ee 
where $\eta$ is the detector efficiency and $P$ is the optical power.

Short laser pulses of $\lambda=532$ nm were used in the experiment
and the external magnetic field was $5T$ \cite{Chou:2007zzc}. The
weakly-interacting ALP interpretation of the PVLAS
data was excluded at more than $5\sigma$ by the GammeV data for
scalar particles. No events were found above the background and
thus a bound was set for the coupling \cite{Chou:2007zzc}: 
\be g\leq 3.1 \times 10^{-7} GeV^{-1}\ee 
This limit is the mean value of two configurations for the magnetic
field and it is valid for small values of the mass
$m_\phi$ (bellow $meV$). Generally, the coupling depends on the
mass of the scalar particle, but when the mass is small (bellow
few $meV$), the coupling is almost unchanged.

 \subsection{ Fifth force experiments}

The coupling between scalar particle and two photons
$\phi\gamma\gamma$ \cite{Dupays:2006dp}, which can be described
with the Lagrangian term (\ref{lagra}), leads to the existence of
long-range non-Newtonian forces (fifth force).
They are bounded by E\"{o}tv\"{o}s type experiments and they don't violate the Equivalence Principle. The
relative difference between inertial and gravity mass is less
than $10^{-12}$ \cite{Su:1994gu} and drives to constrains on the
coupling constant.

The Lagrangian contains a interaction term of the form: 
\be L_{interaction}=-\frac{g\phi}{4}F_{\mu \nu}F^{\mu\nu}-L_2\ee 
The above term of the Lagrangian density induces radiatively a
coupling to charged particles, such as electrons or protons. The
additional term in Lagrangian density is 
$ L_2=y \phi\overline{\Psi}\Psi$ where $y$ is the Yukawa coupling and
$\Psi$ is the field of the charged particle. The authors of
\cite{Dupays:2006dp} used existing experimental limits to
constrain the coupling constant $g$ as a function of the mass of
the scalar field $m_\phi$. These limits emerge from a
micromechanical resonator which measures the  Casimir force
between parallel plates \cite{Perivolaropoulos:2008pg,Bressi:2002fr} (two mirrors in
a vacuum will be attracted to each other) placed a few nanometers
apart, from experiments with torsion pendulum and a rotating
attractor \cite{Hoyle:2004cw} and from experiments which use
torsion-balance \cite{Adelberger:2006dh}. Using the last class of
experiments, the authors \cite{Dupays:2006dp} reached very
stringent results when the field satisfies the condition
$\Lambda\gg m_p$ ($\Lambda$ is the cosmological constant and
$m_p$ the proton-mass). When $m_\phi\sim meV$, they found that \cite{Adelberger:2006dh}:  
\be g< 1.6 \times 10^{-17} GeV^{-1}\ee
It is a stringent limit and the terrestrial  experiments don't have until now, the sensitivity to detect some event.

Scalar particles with almost zero mass can lead not only to long-range forces (in the same manner as quintessence),
but also to variation of fundamental constants \cite{Lee:2004vm}. Bekenstein type models with a scalar field
$\phi$, that affects the electromagnetic permeability, lead to variations of the effective fine structure constant up to very
high red-shifts. The coupling between scalar field and electromagnetic tensor of the form:
\be\beta_{F^2}(\phi/M)F_{\mu \nu} F^{\mu\nu}\equiv \frac{g}{4} \phi F_{\mu \nu} F^{\mu\nu}\ee
can lead to a time variation \cite{Carroll:1998zi} of the fine structure constant, due to the time variation of the scalar field. The scalar field $\phi$ is expected to have a variation at the present time (in cosmological timescales) of order $M_{Pl}$ and there are several observations to bound such variation. From the Oklo natural reactor in Gabon \cite{Landau:2000cc}, the researchers analyzed the isotope ratios
of $^{149}Sm/^{147}Sm$ in the natural uranium fission reactor
(mine) that operated $1.8$ billion years ago. The isotopic
abundances lead to $|\dot{\alpha}/\alpha|<10^{-15} yr^{-1}$ over
the last $1.8$ billion years and constrains the coupling as:
\be g\leq 4\times10^{-6}(\frac{H_0}{\langle\dot{\phi}\rangle})\ee
where $H_0 \sim 10^{-33} eV$ and $\langle\dot{\phi}\rangle$ is the
mean rate of change of $\phi$ in the above time range.

\subsection{ BFRT experiment}

One of the first photon regeneration experiments took place in
Brookhaven National Laboratory \cite{Ruoso:1992nx}. In this experiment the beam had wavelength $\lambda=514 nm$ and
the magnetic field was $3.7T$. The search for scalar particles
requires the laser polarization to be perpendicular to the
magnetic field. The photons, produced during the regeneration, are
detected by sensitive photocathode of a photomultiplier tube (PMT)
\cite{Davis:2007wu}. For $220$ minutes the laser was on and
subsequently for $220$ minutes the laser was off. They didn't
observe significant difference between laser on and laser off
states. Thus \cite{Ruoso:1992nx}, \cite{Jaeckel:2006xm}, in the
absence of signal, the coupling constant was constrained as: 
\be g < 6.7 \times 10^{-7} GeV^{-1}\ee 
at $90 \% $ confidence level. This limit is applicable when the scalar particle is very
light with mass $m < 10^{-3} eV$. The PVLAS signal and the BFRT constraint can be combined as \cite{Jaeckel:2006xm}: 
\be 1.7 \times 10^{-6} GeV^{-1} \leq g \leq 5 \times 10^{-6} GeV^{-1} \ee 
assuming the mass of the scalar in the range $1 meV \leq m_\phi \leq 1.5 meV$.

\subsection{ OSCAR experiment}

The OSCAR experiment takes place at LHC and it is a photon regeneration experiment which uses two LHC dipole magnets.
The laser beam has wavelength $\lambda=514 nm$ and the dipole superconducting magnets are cooled down to $1.9K$ \cite{Pugnat:2007nu}. 
The innovation in this experiment is that they use a buffer of neutral gas as a resonant amplifier medium. The conversion probability, divided by
the refractive index $n=\sqrt{\varepsilon}$, is: 
\be P_{\gamma\leftrightarrow s}= \frac{1}{4u\sqrt{\varepsilon}}(g B L)^2(\frac{2}{q L}\sin\frac{q L}{2})^2\ee 
while the expected counting rate is given by equation (\ref{rate}). The device of the OSCAR 
experiment hasn't recorded any signal and the coupling is constrained as \cite{Schott:2011fm}: 
\be g < 1.15 \times 10^{-7} GeV^{-1}\ee

An updated result \cite{Ballou:2014myz}, is currently the lowest limit from such experiments. In the case of massless scalar 
particle, the coupling constrained as: 
\be g < 5.76 \times 10^{-8} GeV^{-1}\ee 
at $95\%$ confidence limit.

\subsection{ ALPS experiment}

The ALPS (Any Light Particle Search) is another one experiment, which based
on the effect "light shining through the wall". The experiment takes place in Deutsches
Electronen Synchrotron (DESY), in Germany \cite{Ehret:2009sq,Ehret:2010ki}. The researchers use a HERA superconducting
dipole magnet where the magnetic field is $5T$. The photons have wavelength $\lambda=1024$nm, or
$\lambda=512$nm. They collect data in vacuum and in low pressure gas, inside a tube, but in the absence of any positive signal for photon
regeneration, they estimated \cite{Ehret:2010mh} the coupling constant as: 
\be g < 7 \times 10^{-8} GeV^{-1}\ee  
in the case of massless scalar particle in vacuum.

\subsection{ LIPSS experiment}
The Light Pseudoscalar and Scalar Particle Search (LIPSS) collaboration \cite{Afanasev:2008jt} was another similar
experiment, looked for photons coupled to light neutral particles. It took place in Jefferson Lab in the Spring of $2007$ and was also based on the light shining through the wall effect. The magnetic field was $1.77T$ for both generation and regeneration regions. The wall was a mirror and the wavelength
of the photons was $\lambda=935 nm$. The innovation of this approach was that data were taken for longer time (almost $1$
hour), than previous similar experiments. No signal was recorded above background and the constraint \cite{Afanasev:2008jt} on the coupling strength is: 
\be g< 10^{-6} GeV^{-1}\ee 
assuming a mass of the scalar particle of order $meV$.

\begin{table}[t]
\centering \scalebox{0.95}{
\begin{tabular} {|c|c|c|c|}
 \hline
experiment & g($ \times GeV^{-1}$)& $m_{\phi}$ & effect\\
  \hline
  PVLAS \cite{Chou:2007zzc}& $\sim 2.5\times 10^{-6} $ & $\sim 1.2meV$ & birefrigence\\
  GammeV \cite{Chou:2007zzc}& $\leq 3.1 \times 10^{-7} $ & $\leq 1meV$ & LSW \\
  Fifth force \cite{Adelberger:2006dh}& $< 1.6 \times 10^{-17} $ & $\sim meV $  & Casimir force \\
  BFRT \cite{Jaeckel:2006xm}& $< 6.7 \times 10^{-7}$ & $\leq 1meV$ & LSW\\
  OSCAR \cite{Ballou:2014myz}& $< 5.76 \times 10^{-8} $& massless & LSW\\
  ALPS \cite{Ehret:2010mh}& $< 7 \times 10^{-8} $& massless & LSW\\
  LIPSS \cite{Afanasev:2008jt}& $< 1 \times 10^{-6} $ & $ \sim meV$ & LSW\\
  \hline
\end{tabular}
} \caption{Constrains on coupling between photons and
scalar particles from all known experiments. Each limit is valid for the
corresponding range of the mass of the scalar particle, which
is shown in the third column. In the fourth column we show the basic
physical effect on which each experiment is based (LSW means light shinning through a wall). } \label{tab}
\end{table}

In table \ref{tab} we present the constrains on the coupling between scalar particles and photons, from all known
experiments in order to compare them and identify the most stringent. For small masses of the scalar particle (bellow $meV$),
the coupling $g$ is mass independent, because the oscillation length between ALPs and photons far exceeds the
length of the magnet. As we see, the controversial result of PVLAS leads to a weak constraint. The other experiments give more
stringent bounds, which in fact aren't consistent with the PVLAS result.

\begin{figure}[!ht]
\centering
\includegraphics[scale=0.6]{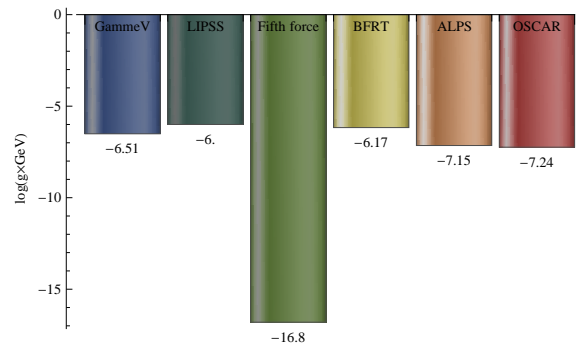}
\caption{Constrains on scalar coupling to electromagnetism from
all known experiments, except PVLAS experiment. The vertical axis
shows the common logarithm of the coupling $g$, dimensionless.}
\label{g2}
\end{figure}

Also, the data of table \ref{tab} are shown through a histogram in Figure \ref{g2}. We have neglected the PVLAS
experiment, because the Italian collaboration doesn't defend it. Thus, the most stringent bound obtained from the
fifth force experiments.

\section{ Constraints on Chameleons}

The existence of chameleons \cite{Brax:2004qh,Nojiri:2003ti,Brax:2006np,Brax:2007vm} could support the accelerating expansion of the universe (as
components of dark energy) and the time evolution of the fine
structure constant. Chameleons are scalar particles
\cite{Nelson:2008uy,Brax:2007ak} whose effective mass is a
function of its local environment. Just like a chameleon changes
color in different environments, the magnitude of the mass of a cosmological chameleon particle
depends on the location. In regions with high density, such as
Earth, the mass is large in order to evade the fifth force
searches, which excluded by experiments on a wide range of scales.
In regions with low density, such as our solar system, the mass is
lower and in cosmological scales is of order of the
present Hubble value \cite{Hinterbichler:2010wu,Creminelli:2013nua}. In any case the effective mass is: 
\be m_{eff}\equiv \sqrt{\frac{d^2V_{eff}}{d\phi^2}}\label{mcha}\ee 
with 
\be V_{eff}(\phi,\vec{x})=V(\phi)+e^{\frac{\beta_m\phi}{M_{Pl}}}\rho_m(\vec{x})+e^{\frac{\beta_\gamma \phi}{M_{Pl}}}\rho_\gamma(\vec{x})\ee 

It is clear, that the effective mass depends on the electromagnetic fields and the local
matter density \cite{Joyce:2014kja}. A possible potential for chameleons is the Ratra-Peebles potential: 
\be V(\phi)= M^4(\frac{M}{\phi})^n\ee 
with $n$ is an integer and $M$ is model parameter (in the case of dark matter, $M\simeq 3meV$). When the
local matter density is high, the chameleon becomes invisible due to mixing with the environment. For this reason
experiments, which have the purpose to detect chameleons in laboratory, are performed in almost absolute vacuum.

Chameleons can couple to all forms of matter and can also couple to photons \cite{Schelpe:2010he}. Coupling to matter leads to fifth force which act only on large scales and is very small on small scales. We do not see any fifth force or modification of gravity in the laboratory or in the Solar System.  The chameleon mechanism has exactly the above  properties, because it suppresses the fifth force mediated by the new degree of freedom without killing the modification on all scales. The environment dependent mass (\ref{mcha}) is enough to hide the fifth force in dense media such as the atmosphere. The chameleon force \cite{Khoury:2003aq,Khoury:2003rn} is only sourced by a thin shell near the surface of dense objects, which reduces its magnitude significantly. 

Chameleon theories are intriguing and lead to new physics. Hints of such theories have been seen in active galactic nuclei's (AGNs) and in the structure of starlight polarization. In conclusion, chameleon mechanism make this class of models cosmologically interesting despite of the strong laboratory constrains imposed by the fifth force experiments.

There are two classes of experiments for searching chameleons:
\begin{itemize}
    \item experiments in empty, closed \textbf{container or jar}, such as GammeV and CHASE.
    \item experiments in microwave \textbf{cavity}, such as ADMX.
\end{itemize}

They are based on the coupling between photons and chameleons, where the coupling to electromagnetism is dominant. These
experiments aren't photon regeneration experiments \cite{Chou:2008gr} because the mass of the chameleons depends on
local density and thus they can't pass through the wall. Inside the wall the density is high compared to the vacuum and the
chameleons get reflected by the wall. These experiments are based on the afterglow effect \cite{Ahlers:2007st}, which we
describe below. In both cases the coupling to electromagnetism may be described by a dilatonic function: 
\be B_F(\phi)= e^{g\phi}\simeq 1+g\phi\ee
because $g$ is very small ($g\ll 1$). This coupling allows photon-chameleon oscillations in the presence of an external
strong magnetic field. The scalar field $\phi$ with mass $m_\phi$ expected of order $meV$. Such a mass could explain
the dark energy density, which is $\sim (meV)^4$.

\subsection{ GammeV experiment}

The GammeV collaboration includes experiments for ALPs and experiments for chameleons
\cite{Chou:2008gr}, \cite{Steffen:2008au}, which couple
to photons. It constitutes the first test of dark energy
models in laboratory. Chameleons produced
inside a optical transparent jar from photon oscillations
(Primakoff effect \cite{Brax:2009bk}) and trapped there, if its
total energy is less than its effective mass. Then, the chameleons
reflected by the walls and they detected via their afterglow as
they slowly converted to photons. The afterglow is possible if the
mixing time between scalars and photons is larger than the
travelling time of photons into the chamber. An afterglow photon
can be observed by a photomultiplier tube (PMT) at the exit
window, when the original photon source (laser) is tuned off. The
pressure in the chamber is $P\approx10^{-7} Torr$ and the
probability per photon to chameleon production is: 
\be P_{pr}=\frac{4g^2B^2\omega^2}{m_{eff}^4}sin^2(\frac{m_{eff}^2L}{4\omega})\widehat{k}\times(\widehat{x}\times\widehat{k})\ee
proportional to the square of coupling $g$.

The magnetic field is in the $\widehat{x}$ direction and $\widehat{k}$ is the direction of motion of the particle. It is clear that, if we
want to have the maximum probability, the photons must propagate in a direction perpendicular to magnetic field.  The photons
have energy $2.33eV$, production rate $\sim10^{19}$ photons per second and the magnetic field is $5T$.

In this case, the action which describes the coupling between photons and chameleons is: 
\begin{eqnarray} S &=& \int d^4x (-\half\,\partial_\mu\,\phi\partial^\mu\,\phi -V(\phi)-\frac{e^{\phi/M_\gamma}}{4} F_{\mu \nu} F^{\mu\nu}\nonumber\\&+&\mathcal{L}_m(e^{2\phi/M_m}g_{\mu\nu},\psi_{m}^{\imath}))\label{action1} \end{eqnarray} 
where $V(\phi)$ is the chameleon potential and $\mathcal{L}_m$ the Lagrangian density for the matter. The
coupling to matter defined as $\beta_m=M_{Pl}/M_m$ and the coupling to electromagnetism is the
dimensionless parameter $\beta_\gamma=M_{Pl}/M_\gamma\equiv g M_{Pl}$. Data were taken for one hour after the laser turned off,
but there wasn't detection of any significant signal in the highly sensitive PMT. Thus, the parameter $g$ estimated as: \cite{Chou:2008gr} 
\be 2.1\times10^{-7}GeV^{-1}< g <2.7\times10^{-6}GeV^{-1}\label{e2}\ee
This limit valid  for coherent oscillations and therefore the effective mass must be quite small ($m_{eff}\ll 0.98 meV$).

\subsection{ ADMX experiment}
The Axion Dark Matter experiment has two parts. The first is the
search for ALPs and the second, the search for
chameleons. In both cases the particles interact with photons
inside a cavity and estimated the range of the coupling. 
The advantage of the microwave cavity is, that the resonance
is stronger than the case where laser is used. This effect
increases the conversion probability and the expected counting
rate of photons in the detector. A microwave receiver amplifies
the excitation of the resonance. The mixing is maximum when
photons and chameleons have the same energy
($\omega_{cham.}=\omega_\gamma$). It is crucial to emphasize that
if the coupling is very weak, the chameleons don't have enough
energy to be detected, while if the coupling is very strong the
chameleons immediately decay.

As discussed in \cite{Rybka:2010ah}, this experiment used a magnet
$7T$, while the cavity had volume $220\ell$. It was hold under
vacuum at $2$ Kelvin. No significant signal was observed and the
excluded region was estimated as \cite{Rybka:2010ah}: 
\be 3.75\times10^{-9}GeV^{-1}< g <2.1\times10^{-4}GeV^{-1}\ee 
at $90\%$ confidence level. This bound is valid for a very small
range of the effective mass, between $1.9510 \mu eV$ and $1.9525
\mu eV$. The above limit overlaps with the limit (\ref{e2}).

\subsection{ CHASE experiment}

The Chameleon Afterglow Search Experiment (CHASE) is a continuation of the GammeV experiment in the same laboratory \cite{Steffen:2010ep}. The excluded region for the chameleon-photon coupling in this case, is significantly improved. Also, the results smooth out the differences between the two
previous experiments \cite{Upadhye:2009iv}.

The novelty of this experiment is twofold. First, it uses two
glasses into the cavity. Thus, the magnetic field is divided in
three parts with different ranges. The shorter part has
sensitivity to chameleons with high mass. Second, in order to
improve the sensitivity for large $g$, used several magnetic
fields, with values lower than $5T$. Finally, in order to improve
the sensitivity for small $g$, a shutter (chopper) is used to
modulate any possible signal from afterglow. The data didn't show any signal of a photon-chameleon coupling
and the excluded region for $m_{eff}\leq 1meV$ is estimated as \cite{Steffen:2010ep}: 
\be 4\times10^{-6}GeV^{-1}< g <1.3\times10^{-3}GeV^{-1}\ee 
at $90\%$ confidence level.

\begin{table}[t]
\centering \scalebox{0.9}{
\begin{tabular}{|c|c|c|}
  \hline
  experiment & excluded $g (\times GeV^{-1})$ & $m_{eff}$ \\
  \hline
  GammeV \cite{Chou:2008gr} & $ (2.1\times10^{-7},2.7\times10^{-6}$) & $\ll 0.98meV $ \\
  ADMX \cite{Rybka:2010ah} & $ (3.75\times10^{-9},2.1\times10^{-4})$ & $[1.9510\mu eV,1.9525\mu eV] $ \\
  CHASE \cite{Steffen:2010ep} & $(4\times10^{-6},1.3\times10^{-3})$ & $\leq 1meV $\\
  \hline
\end{tabular}
}
 \caption{Excluded regions on coupling between photons and
chameleons from all known afterglow experiments. In third column,
recorded the corresponding effective mass for the chameleons.}
\label{tab:gcha}
\end{table}

\section{ Cosmological and Astrophysical Effects}
We can extend the Bekenstein theory when we introduce the dilatonic function $B_F(\phi)=e^{-2\phi}$ in formula (\ref{glagra}). This function induces effects on the cosmological evolution of a quintessence scalar field \cite{Barrow:2013uza} and effects of multidimensional gravity \cite{Bronnikov:2013xh}. We discuss these effects in some detail. 

We want to investigate the cosmological evolution and the effect of the new coupling on the Big Crunch singularity \cite{Perivolaropoulos:2004yr,Kallosh:2003bq,Lykkas:2015kls} that is present in linear potentials. In Ref. \cite{Barrow:2013uza} the authors introduced the Lagrangian density  
\be{L}=\frac{R}{2}-\frac{\omega(\phi)}{2}\partial_{a}\phi \partial^{a}\phi-V(\phi)-\frac{1}{4}e^{-2\phi} F_{\mu\nu} F^{\mu \nu}+L_m\label{lqui}\ee 
where the fine structure constant varies through the relation $\alpha=\alpha_{0}e^{-2\phi}$. We consider FRW flat spacetime, with $\omega(\phi)=1$ and $V(\phi)=-s\phi$. We introduce the rescaling $H=\bar{H}H_0$, $t=\frac{\bar{t}}{H_0}$, $V=\bar{V}H_0^2$, $\rho_m=\bar{\rho_m}H_0^2$ and $\rho_r=\bar{\rho_r}H_0^2$, ($H_0$ is the present value of the Hubble constant) in the dynamic equations of Ref \cite{Barrow:2013uza} and from now on we omit the bar.  
Thus, the scalar field equation of motion takes the form 
\be \ddot{\phi}+3H\dot{\phi}+V'(\phi)=\frac{-6\zeta_m \Omega_{0m}e^{-2\phi} }{a^3(1+|\zeta_m|e^{-2\phi_0})}\label{eqphi} \ee 
where $\phi_0$ is the present value of the scalar field and $\zeta_m=L_{em}/\rho_m$. Here, $\rho_m$ is the energy density of non-relativistic matter. In a radiation epoch, variations in fine structure constant are driven only by the electromagnetic energy of non-relativistic matter, because $L_{em}=\frac{1}{2}(E^2-B^2)=0$. 

Respectively, the acceleration equation for the scale factor becomes \cite{Barrow:2013uza}  
\be \frac{\ddot{a}}{a}=-\frac{\Omega_{0m}(1+|\zeta_m|e^{-2\phi})}{2a^3(1+|\zeta_m|e^{-2\phi_0})}-\frac{\Omega_{0r}e^{-2(\phi-\phi_0)}}{a^4}-\frac{1}{3}[\dot{\phi}^2-V(\phi)]\label{eqa}\ee

\begin{figure}[!ht]
\centering
\includegraphics[scale=0.65]{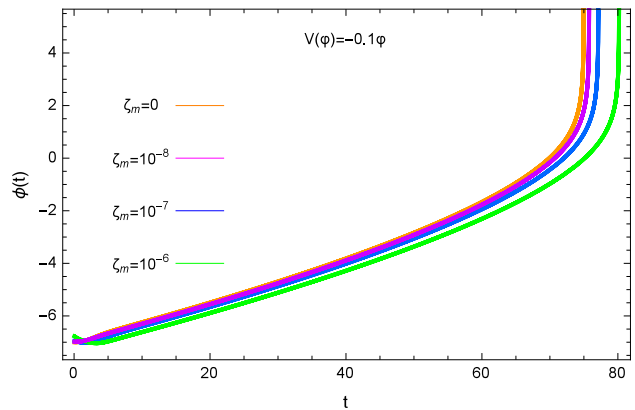}
\caption{The scalar field $\phi(t)$ as a function of time $t$ when $\zeta_m=0$, $\zeta_m=10^{-8}$, $\zeta_m=10^{-7}$ and $\zeta_m=10^{-6}$, when the potential is of the form $V(\phi)=-0.1\phi$. The present time $t_0$ is derived from the solution and must be almost equal to $1$. As we see, the field after some time increases quickly, thus the effective force becomes attractive.}
\label{g3}
\end{figure}

\begin{figure}[!ht]
\centering
\includegraphics[scale=0.65]{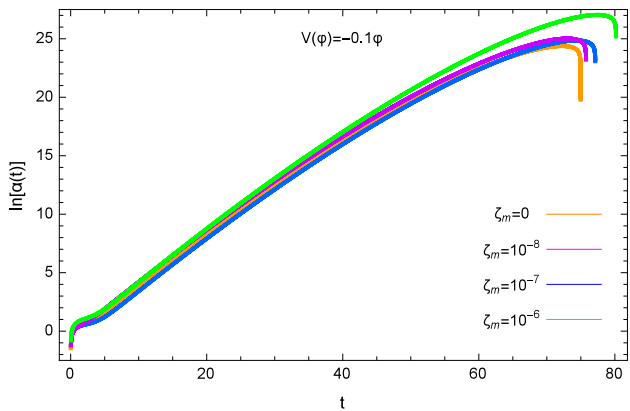}
\caption{The common logarithm of the scale factor $ln(a(t))$ as a function of time $t$, for several values of the parameter $\zeta_m$, when the potential is of the form $V(\phi)=-0.1\phi$ and $t_0\simeq1$. After some time the system begins to shrink due to attractive force, caught by the scalar field. As $\zeta_m$ increases the Big Crunch occurs later.}
\label{g4}
\end{figure}

We have solved the system of the cosmological dynamical equations for the scalar field and for the scale
factor (\ref{eqphi}) and (\ref{eqa}). We assume $\Omega_{0r}=10^{-4}$, $\Omega_{0m}=0.3$ and initial conditions deep in the radiation era where the scalar field $\phi_i$ was almost constant ($\dot{\phi}(t_i)=0$). Due to rescaling, the acceptable solutions must satisfy the conditions $a(t_0)=1$, $H(t_0)=1$ and $\Omega_{0\phi}=0.7$, where $t_0$ is the present time. In fig. \ref{g3} we present the scalar field as a function of time when $V(\phi)=-0.1\phi$, while in fig. \ref{g4} we have plot the logarithm of the scale factor $ln(a(t))$.

It is clear that, when the scalar field increases rapidly, the effective force becomes attractive, the scale factor decreases also rapidly and the Universe leading to Big Crunch. When $\zeta_m$ increases, the effect occurs later. This is an expected result, if we carefully observe the equation (\ref{eqphi}). The right hand side is a function of $\zeta_m$ and as $\zeta_m$ increases, the r.h.s decreases (bellow zero). Thus, the scalar field needs more time to begin increasing rapidly. In other words, the presence of non-relativistic matter stabilises the system for longer time (before Big Crunch).

\begin{figure}[!ht]
\centering
\includegraphics[scale=0.65]{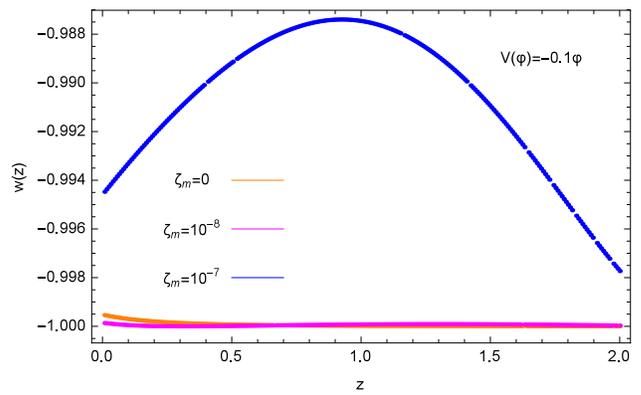}
\caption{The equation of state parameter $w(z)$ as a function of redshift $z$ when $\zeta_m=0$, $\zeta_m=10^{-8}$, and $\zeta_m=10^{-7}$ when the potential is of the form $V(\phi)=-0.1\phi$ and $t_0\simeq1$. As we see the model describes quintessence cosmology ($w>-1$), but it is close to a cosmological constant $w=-1$.}
\label{g1}
\end{figure}

\begin{figure}[!ht]
\centering
\includegraphics[scale=0.65]{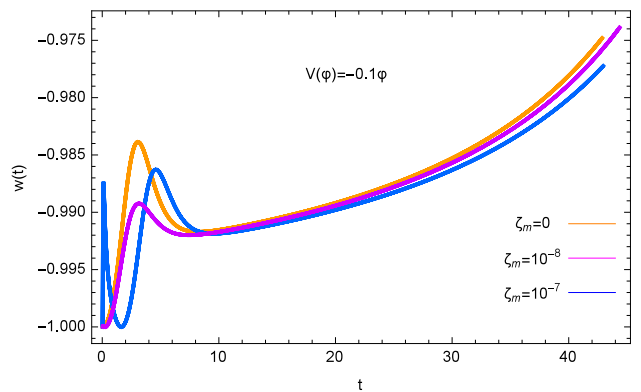}
\caption{The equation of state parameter $w(t)$ as a function of time $t$ for several values of the parameter $\zeta_m$, when the potential is of the form $V(\phi)=-0.1\phi$ and $t_0\simeq1$. }
\label{g5}
\end{figure}

Then, using the solution, we calculated the scalar field dark energy (DE) equation of state parameter $w_{DE}=\frac{P_{DE}}{\rho_{DE}}$ as a function of redshift $z$ through the relation 
\be w(z)=\frac{0.5\dot{\phi}^2+V(\phi)}{0.5\dot{\phi}^2-V(\phi)}\label{eqw}\ee 
and we have plot the results in fig. \ref{g1}. The model corresponds to quintessence cosmology because $w>-1$ and as we see (magenta or green line), the dilatonic function induces small changes in the parameter $w(z)$, if we compare with the case $\zeta_m=0$ (red line). Specifically, when the parameter $\zeta_m$ increases, the equation of state parameter $w(z)$ also increases in the context of quintessence cosmology. Also, in fig \ref{g5}, we have plot the parameter $w$, as a function of time.

The dilatonic function $B_F(\phi)=e^{-2\phi}$ can also describe spatial variations of fine structure constant in nonlinear multidimensional theories of gravity \cite{Bronnikov:2013xh,Bronnikov:2009ai,Bronnikov:2003rf}. This term arises naturally from the metric determinant, by taking into account spatial perturbations (of order of the cosmological horizon scale) of the scalar field and the metric, when the system reduces to four dimensions. The observational data of variations of $\alpha$ depend on the size of the extra factor space and define the model parameters. In this cosmological model, the values of fine structure constant changes slightly or remain almost constant in all cosmological epochs (radiation epoch, matter epoch or accelerating expansion epoch due to a cosmological constant). This process can be used for the research of variations and other fundamental constants, such as the gravitational constant $G$ \cite{Bronnikov:2001th}.    

Large scale inhomogeneity of the scalar field $\phi$ of multidimensional origin can induce spatial variations of $\alpha$. The variations of $\alpha$ are very small (of order $10^{-6}$), as we have mentioned in the introduction \cite{Webb:2010hc} and have been observed from Very Large Telescope (VLT) in Chile \cite{Pettini:2001wp} and Keck telescope in Hawaii \cite{Vogt:1995zz,Faber:2003zz}. The results obtained from spectra of distant quasars and shows a smaller value for fine structure constant when $z<1.8$ from both telescopes. When $z>1.8$, the Keck data shows that $\frac{\Delta\alpha}{\alpha}<0$, but the VLT data drives to $\frac{\Delta\alpha}{\alpha}>0$. The combined dataset fits a spatial dipole for the variation of $\alpha$, which is unlikely to be caused by systematic effects.

There are many cosmological and astrophysical observations, which
could be explained by the existence of scalar ALPs or chameleons and their coupling with photons. One of them is the
dark energy density of the universe \cite{Weber:2009pt,Asztalos:2001jk,deBoer:2010eh}, which is of
order $\rho_\Lambda \sim (meV)^4$. If the scalar ALPs or chameleons exists and have masses of order of $meV$,
the vacuum energy density has the cosmologically required value.

Scalar dark radiation with a sector $\phi$ of spin-$0$, can be
tightly coupled to thermal plasma of hydrogen, $\alpha$ particles,
baryons, photons and electrons. Such a  particle can be scattered from the
plasma. The full Lagrangian \cite{Marsh:2014gca} in this case has the form: 
\be L_{total}=L_{vissible}+L_{dark matter}+L_{interaction}\ee 
where $L_{int.}$ contains the coupling between ALPs and plasma. This term includes
Yukawa-type and dilaton-like operators and has the form \cite{Marsh:2014gca}: 
\be L_{interaction}=-\frac{g\phi}{4}F_{\mu \nu}F^{\mu\nu}-\sum_{i}^{}\frac{m_i}{\Lambda_{4i}}\phi\overline{\psi}_i\psi_i\ee
Astronomical observations \cite{Jaeckel:2006xm}, \cite{Giannotti:2015kwo} from the duration
of the red giant phase and the population of Helium Burning stars
(helium burning generates enough energy to prevent further
contraction of the star core) in globular clusters \cite{Beringer:1900zz}, require:  
\be g<6.25\times10^{-11}GeV^{-1}\ee

This constraint isn't as stringent as experimental constraints due
two uncertainty effects. First, the ALPs may be emitted with less  energy than produced, due to
stellar medium diffusion and second, there may be much
less ALPs produced due to a possible stellar suppression
mechanism. Scattering rate of scalar dark radiation near the above
bound of $g$, will be too small to significantly distort the CMB
blackbody spectrum. Stronger limits on $g$ can be extracted by
considering the cosmological evolution of the vacuum expectation
value of $\phi$.

There are many cosmological sources, such as quasars
\cite{Payez:2011sh}, X-rays from the Sun \cite{Zioutas:2006ns},
cosmic rays with ultra high energy (of order $10^{18} eV$)
\cite{Gorbunov:2001gc}, which produce photons. These photons can
be converted to scalar particles due to magnetic fields, around
their sources. They travel to Earth and reconvert back to
photons due to magnetic fields from our galaxy, or due to
intergalactic or intracluster magnetic fields ('cosmic photon regeneration'). The photons
can be detected through experiments on Earth. The required mass
\cite{Burrage:2009mj} for the ALPs situated in the range
$ m_{a}\ll (1peV - 1neV$) and the required coupling is: 
\be g\sim (10^{-12} - 10^{-11}) GeV^{-1}\ee 

The existence of such particles could explain the alignment of the polarization from distant
quasars \cite{Payez:2008pm}, the variations in luminosity of
active galactic nuclei \cite{Burrage:2009mj}, the Sun activity in
X-rays \cite{Zioutas:2009bw}, the unexpected existence of ultra
high energy cosmic rays \cite{Fairbairn:2009zi} and the detection
of $TeV$ gamma rays \cite{Hochmuth:2007hk} from very distant
cosmological sources on Earth (usually they absorbed high).

Scalar particles can also be produced inside the stars
\cite{Brax:2010gp} and their properties depend on the density of
the environment \cite{Jaeckel:2006xm}. They can be produced in
stellar plasma, only if their mass is tuned to be resonant with
the frequency of the plasma \cite{Brax:2010xq}.

Also, scalar fields can change the energy of the bound states in
atoms \cite{Brax:2010gp}. The nuclear electric field, in and
around the atom, induce a perturbation to scalar field and the
corresponding energy levels of hydrogenic atoms are shifted. Thus,
the gap between the energy levels increases. These shifts (for example
Lamb shift), can be used to constrain the parameter $g$.  The
energy gap between the levels $2S_{1/2}$-$2P_{1/2}$ requires
$g\leq10^{-3} GeV^{-1}$, so it is easier to detect scalar
couplings in laboratory experiments from photon regeneration
experiments than from atomic measurements \cite{Jaeckel:2006xm}.

ALPs maybe emitted by explosion of Supernovae.
They could be produced by the Primakoff effect with energy $E\sim
100MeV$ and finally converted into high energy photos in the
magnetic field of our Galaxy. For example, at a distance
$50kpc$ of Milky Way is the remnant of $SN1987A$, in Large
Magellanic Cloud. The authors of Ref. \cite{Payez:2014xsa} used
the current models for the Supernova magnetic field and the Milky
Way magnetic field and they obtained a bound for the coupling
between photons and ALPs. In the future, any
supernova core-collapse could be used to detect this process.

The coupling between photons and chameleons can also be observed
through effects in light from astrophysical sources
\cite{Burrage:2008ii}. This coupling can induce linear and
circular polarization which can be detected on Earth. The
intergalactic region has very low density \cite{Olive:2007aj},
where the chameleons behave as ALPs. They must
have mass $m_\phi\lesssim 10^{-11}eV$, the range of the chameleon
force is $\lambda_\phi\gtrsim 20Km$ and the required coupling is $
g\gtrsim10^{-11} GeV^{-1}$.

The dilatonic function $B_F(\phi)=1+g(\phi-\phi_0)$ can describe variations of the
fine structure constant \cite{Olive:2007aj}. The evolution of
$\alpha$ is given by the relation \cite{Menezes:2005tp} 
$\frac{\Delta\alpha}{\alpha}=(B_F(\phi))^{-1}-1$. Assuming that $\phi_0=0$,
scalar particles or chameleons would change the value of this
constant, when they interact with photons. If we determine the
order of coupling $g$, then we would check if this value could
support the observed variation of the fine structure constant.

\section{ Conclusions}

The existence of scalar (or pseudoscalar) ALPs and
chameleons can play the role of dark matter or can induce the accelerating expansion of the universe.
The detection of these light particles (with masses
in the sub-eV range) is a very difficult problem
\cite{Baker:2013zta}. For this purpose, many experiments until today have been
designed and executed. There are laboratory experiments and astrophysical or cosmological
observations based on light shining through the wall effect, optical
effects in laser polarization, etc \cite{Ringwald:2014vqa} in order to detect the coupling
between scalar particles and photons through the effects, that induce in light. These experiments
haven't recorded any positive signal, because the coupling, as it
seems from the results, is very weak.

We examined the case where the coupling, described by a dilatonic
function, varies linearly with the scalar field $\phi$ (\ref{bf}). Due to the shift
symmetry of scalar field, quadratic terms of $\phi$ are excluded.
Experiments are being conducted, which try to detect optical
effects from these particles in polarized laser beam, or photon
regeneration inside a strong magnetic field. The detection
sensitivity of these experiments is restricted by the technical
features of each apparatus. In these experiments there isn't
currently any positive signal for the existence of ALPs, so we currently have an upper bound. The most stringent bound
comes from the fifth force experiments where long range forces are induced by the scalar field (Casimir force).

An alternative way to explain the accelerating expansion of the universe is the chameleon scalar particles, a kind of
particles whose mass depends on the local density. In dense environment, the chameleon becomes massive (mediate a short range force), but in sparse environment becomes very light (mediate a long range force) \cite{Hamilton:2015zga}. This feature makes chameleons consistent with local experiments but still effective on the cosmological dynamics beyond the cosmological constant. They are coupled to photons and this coupling can
described with the same dilatonic function as scalar ALPs. The experiments are based on different effects because
the chameleons get reflected by the wall, due to their mass, so they cannot induce photon regeneration.

In many astrophysical, astronomical and cosmological effects, light travel from one distant source, to our
planet. It passes through several magnetic fields and it is possible to detect changes in light, when we observe it, in laboratory.

\section{ACKNOWLEDGMENTS}

The author would like to thank Professor L. Perivolaropoulos for
stimulating discussions on the ALPs, chameleons and
their coupling to electromagnetism, whose feedback improved the
quality of the paper.
 
\bibliography{alp}

\begin{thebibliography}{116}%
\makeatletter
\providecommand \@ifxundefined [1]{%
 \@ifx{#1\undefined}
}%
\providecommand \@ifnum [1]{%
 \ifnum #1\expandafter \@firstoftwo
 \else \expandafter \@secondoftwo
 \fi
}%
\providecommand \@ifx [1]{%
 \ifx #1\expandafter \@firstoftwo
 \else \expandafter \@secondoftwo
 \fi
}%
\providecommand \natexlab [1]{#1}%
\providecommand \enquote  [1]{``#1''}%
\providecommand \bibnamefont  [1]{#1}%
\providecommand \bibfnamefont [1]{#1}%
\providecommand \citenamefont [1]{#1}%
\providecommand \href@noop [0]{\@secondoftwo}%
\providecommand \href [0]{\begingroup \@sanitize@url \@href}%
\providecommand \@href[1]{\@@startlink{#1}\@@href}%
\providecommand \@@href[1]{\endgroup#1\@@endlink}%
\providecommand \@sanitize@url [0]{\catcode `\\12\catcode `\$12\catcode
  `\&12\catcode `\#12\catcode `\^12\catcode `\_12\catcode `\%12\relax}%
\providecommand \@@startlink[1]{}%
\providecommand \@@endlink[0]{}%
\providecommand \url  [0]{\begingroup\@sanitize@url \@url }%
\providecommand \@url [1]{\endgroup\@href {#1}{\urlprefix }}%
\providecommand \urlprefix  [0]{URL }%
\providecommand \Eprint [0]{\href }%
\providecommand \doibase [0]{http://dx.doi.org/}%
\providecommand \selectlanguage [0]{\@gobble}%
\providecommand \bibinfo  [0]{\@secondoftwo}%
\providecommand \bibfield  [0]{\@secondoftwo}%
\providecommand \translation [1]{[#1]}%
\providecommand \BibitemOpen [0]{}%
\providecommand \bibitemStop [0]{}%
\providecommand \bibitemNoStop [0]{.\EOS\space}%
\providecommand \EOS [0]{\spacefactor3000\relax}%
\providecommand \BibitemShut  [1]{\csname bibitem#1\endcsname}%
\let\auto@bib@innerbib\@empty
\bibitem [{\citenamefont {Webb}\ \emph {et~al.}(2011)\citenamefont {Webb},
  \citenamefont {King}, \citenamefont {Murphy}, \citenamefont {Flambaum},
  \citenamefont {Carswell},\ and\ \citenamefont {Bainbridge}}]{Webb:2010hc}%
  \BibitemOpen
  \bibfield  {author} {\bibinfo {author} {\bibfnamefont {J.~K.}\ \bibnamefont
  {Webb}}, \bibinfo {author} {\bibfnamefont {J.~A.}\ \bibnamefont {King}},
  \bibinfo {author} {\bibfnamefont {M.~T.}\ \bibnamefont {Murphy}}, \bibinfo
  {author} {\bibfnamefont {V.~V.}\ \bibnamefont {Flambaum}}, \bibinfo {author}
  {\bibfnamefont {R.~F.}\ \bibnamefont {Carswell}}, \ and\ \bibinfo {author}
  {\bibfnamefont {M.~B.}\ \bibnamefont {Bainbridge}},\ }\bibfield  {title}
  {\enquote {\bibinfo {title} {{Indications of a spatial variation of the fine
  structure constant}},}\ }\href {\doibase 10.1103/PhysRevLett.107.191101}
  {\bibfield  {journal} {\bibinfo  {journal} {Phys. Rev. Lett.}\ }\textbf
  {\bibinfo {volume} {107}},\ \bibinfo {pages} {191101} (\bibinfo {year}
  {2011})},\ \Eprint {http://arxiv.org/abs/1008.3907} {arXiv:1008.3907
  [astro-ph.CO]} \BibitemShut {NoStop}%
\bibitem [{\citenamefont {Chiba}(2011)}]{Chiba:2011bz}%
  \BibitemOpen
  \bibfield  {author} {\bibinfo {author} {\bibfnamefont {Takeshi}\ \bibnamefont
  {Chiba}},\ }\bibfield  {title} {\enquote {\bibinfo {title} {{The Constancy of
  the Constants of Nature: Updates}},}\ }\href {\doibase 10.1143/PTP.126.993}
  {\bibfield  {journal} {\bibinfo  {journal} {Prog. Theor. Phys.}\ }\textbf
  {\bibinfo {volume} {126}},\ \bibinfo {pages} {993--1019} (\bibinfo {year}
  {2011})},\ \Eprint {http://arxiv.org/abs/1111.0092} {arXiv:1111.0092 [gr-qc]}
  \BibitemShut {NoStop}%
\bibitem [{\citenamefont {Murphy}\ \emph {et~al.}(2003)\citenamefont {Murphy},
  \citenamefont {Webb},\ and\ \citenamefont {Flambaum}}]{Murphy:2003hw}%
  \BibitemOpen
  \bibfield  {author} {\bibinfo {author} {\bibfnamefont {Michael~T.}\
  \bibnamefont {Murphy}}, \bibinfo {author} {\bibfnamefont {J.~K.}\
  \bibnamefont {Webb}}, \ and\ \bibinfo {author} {\bibfnamefont {V.~V.}\
  \bibnamefont {Flambaum}},\ }\bibfield  {title} {\enquote {\bibinfo {title}
  {{Further evidence for a variable fine-structure constant from Keck/HIRES QSO
  absorption spectra}},}\ }\href {\doibase 10.1046/j.1365-8711.2003.06970.x}
  {\bibfield  {journal} {\bibinfo  {journal} {Mon. Not. Roy. Astron. Soc.}\
  }\textbf {\bibinfo {volume} {345}},\ \bibinfo {pages} {609} (\bibinfo {year}
  {2003})},\ \Eprint {http://arxiv.org/abs/astro-ph/0306483}
  {arXiv:astro-ph/0306483 [astro-ph]} \BibitemShut {NoStop}%
\bibitem [{\citenamefont {Mota}\ and\ \citenamefont
  {Barrow}(2004)}]{Mota:2003tm}%
  \BibitemOpen
  \bibfield  {author} {\bibinfo {author} {\bibfnamefont {David~F.}\
  \bibnamefont {Mota}}\ and\ \bibinfo {author} {\bibfnamefont {John~D.}\
  \bibnamefont {Barrow}},\ }\bibfield  {title} {\enquote {\bibinfo {title}
  {{Local and global variations of the fine structure constant}},}\ }\href
  {\doibase 10.1111/j.1365-2966.2004.07505.x} {\bibfield  {journal} {\bibinfo
  {journal} {Mon. Not. Roy. Astron. Soc.}\ }\textbf {\bibinfo {volume} {349}},\
  \bibinfo {pages} {291} (\bibinfo {year} {2004})},\ \Eprint
  {http://arxiv.org/abs/astro-ph/0309273} {arXiv:astro-ph/0309273 [astro-ph]}
  \BibitemShut {NoStop}%
\bibitem [{\citenamefont {Reinhold}\ \emph {et~al.}(2006)\citenamefont
  {Reinhold}, \citenamefont {Buning}, \citenamefont {Hollenstein},
  \citenamefont {Ivanchik}, \citenamefont {Petitjean},\ and\ \citenamefont
  {Ubachs}}]{Reinhold:2006zn}%
  \BibitemOpen
  \bibfield  {author} {\bibinfo {author} {\bibfnamefont {E.}~\bibnamefont
  {Reinhold}}, \bibinfo {author} {\bibfnamefont {R.}~\bibnamefont {Buning}},
  \bibinfo {author} {\bibfnamefont {U.}~\bibnamefont {Hollenstein}}, \bibinfo
  {author} {\bibfnamefont {A.}~\bibnamefont {Ivanchik}}, \bibinfo {author}
  {\bibfnamefont {P.}~\bibnamefont {Petitjean}}, \ and\ \bibinfo {author}
  {\bibfnamefont {W.}~\bibnamefont {Ubachs}},\ }\bibfield  {title} {\enquote
  {\bibinfo {title} {{Indication of a Cosmological Variation of the Proton -
  Electron Mass Ratio Based on Laboratory Measurement and Reanalysis of H(2)
  Spectra}},}\ }\href {\doibase 10.1103/PhysRevLett.96.151101} {\bibfield
  {journal} {\bibinfo  {journal} {Phys. Rev. Lett.}\ }\textbf {\bibinfo
  {volume} {96}},\ \bibinfo {pages} {151101} (\bibinfo {year}
  {2006})}\BibitemShut {NoStop}%
\bibitem [{\citenamefont {Fenner}\ \emph {et~al.}(2005)\citenamefont {Fenner},
  \citenamefont {Murphy},\ and\ \citenamefont {Gibson}}]{Fenner:2005sr}%
  \BibitemOpen
  \bibfield  {author} {\bibinfo {author} {\bibfnamefont {Yeshe}\ \bibnamefont
  {Fenner}}, \bibinfo {author} {\bibfnamefont {M.~T.}\ \bibnamefont {Murphy}},
  \ and\ \bibinfo {author} {\bibfnamefont {B.~K.}\ \bibnamefont {Gibson}},\
  }\bibfield  {title} {\enquote {\bibinfo {title} {{On variations in the
  fine-structure constant and limits on AGB pollution of quasar absorption
  systems}},}\ }\href {\doibase 10.1111/j.1365-2966.2005.08781.x} {\bibfield
  {journal} {\bibinfo  {journal} {Mon. Not. Roy. Astron. Soc.}\ }\textbf
  {\bibinfo {volume} {358}},\ \bibinfo {pages} {468--480} (\bibinfo {year}
  {2005})},\ \Eprint {http://arxiv.org/abs/astro-ph/0501168}
  {arXiv:astro-ph/0501168 [astro-ph]} \BibitemShut {NoStop}%
\bibitem [{\citenamefont {Langacker}\ \emph {et~al.}(2002)\citenamefont
  {Langacker}, \citenamefont {Segre},\ and\ \citenamefont
  {Strassler}}]{Langacker:2001td}%
  \BibitemOpen
  \bibfield  {author} {\bibinfo {author} {\bibfnamefont {Paul}\ \bibnamefont
  {Langacker}}, \bibinfo {author} {\bibfnamefont {Gino}\ \bibnamefont {Segre}},
  \ and\ \bibinfo {author} {\bibfnamefont {Matthew~J.}\ \bibnamefont
  {Strassler}},\ }\bibfield  {title} {\enquote {\bibinfo {title} {{Implications
  of gauge unification for time variation of the fine structure constant}},}\
  }\href {\doibase 10.1016/S0370-2693(02)01189-9} {\bibfield  {journal}
  {\bibinfo  {journal} {Phys. Lett.}\ }\textbf {\bibinfo {volume} {B528}},\
  \bibinfo {pages} {121--128} (\bibinfo {year} {2002})},\ \Eprint
  {http://arxiv.org/abs/hep-ph/0112233} {arXiv:hep-ph/0112233 [hep-ph]}
  \BibitemShut {NoStop}%
\bibitem [{\citenamefont {Marra}\ and\ \citenamefont
  {Rosati}(2005)}]{Marra:2005yt}%
  \BibitemOpen
  \bibfield  {author} {\bibinfo {author} {\bibfnamefont {Valerio}\ \bibnamefont
  {Marra}}\ and\ \bibinfo {author} {\bibfnamefont {Francesca}\ \bibnamefont
  {Rosati}},\ }\bibfield  {title} {\enquote {\bibinfo {title} {{Cosmological
  evolution of alpha driven by a general coupling with quintessence}},}\ }\href
  {\doibase 10.1088/1475-7516/2005/05/011} {\bibfield  {journal} {\bibinfo
  {journal} {JCAP}\ }\textbf {\bibinfo {volume} {0505}},\ \bibinfo {pages}
  {011} (\bibinfo {year} {2005})},\ \Eprint
  {http://arxiv.org/abs/astro-ph/0501515} {arXiv:astro-ph/0501515 [astro-ph]}
  \BibitemShut {NoStop}%
\bibitem [{\citenamefont {Nesseris}\ and\ \citenamefont
  {Perivolaropoulos}(2004)}]{Nesseris:2004uj}%
  \BibitemOpen
  \bibfield  {author} {\bibinfo {author} {\bibfnamefont {S.}~\bibnamefont
  {Nesseris}}\ and\ \bibinfo {author} {\bibfnamefont {Leandros}\ \bibnamefont
  {Perivolaropoulos}},\ }\bibfield  {title} {\enquote {\bibinfo {title} {{The
  Fate of bound systems in phantom and quintessence cosmologies}},}\ }\href
  {\doibase 10.1103/PhysRevD.70.123529} {\bibfield  {journal} {\bibinfo
  {journal} {Phys. Rev.}\ }\textbf {\bibinfo {volume} {D70}},\ \bibinfo {pages}
  {123529} (\bibinfo {year} {2004})},\ \Eprint
  {http://arxiv.org/abs/astro-ph/0410309} {arXiv:astro-ph/0410309 [astro-ph]}
  \BibitemShut {NoStop}%
\bibitem [{\citenamefont {Anchordoqui}\ and\ \citenamefont
  {Goldberg}(2003)}]{Anchordoqui:2003ij}%
  \BibitemOpen
  \bibfield  {author} {\bibinfo {author} {\bibfnamefont {Luis}\ \bibnamefont
  {Anchordoqui}}\ and\ \bibinfo {author} {\bibfnamefont {Haim}\ \bibnamefont
  {Goldberg}},\ }\bibfield  {title} {\enquote {\bibinfo {title} {{Time
  variation of the fine structure constant driven by quintessence}},}\ }\href
  {\doibase 10.1103/PhysRevD.68.083513} {\bibfield  {journal} {\bibinfo
  {journal} {Phys. Rev.}\ }\textbf {\bibinfo {volume} {D68}},\ \bibinfo {pages}
  {083513} (\bibinfo {year} {2003})},\ \Eprint
  {http://arxiv.org/abs/hep-ph/0306084} {arXiv:hep-ph/0306084 [hep-ph]}
  \BibitemShut {NoStop}%
\bibitem [{\citenamefont {Tsujikawa}(2013)}]{Tsujikawa:2013fta}%
  \BibitemOpen
  \bibfield  {author} {\bibinfo {author} {\bibfnamefont {Shinji}\ \bibnamefont
  {Tsujikawa}},\ }\bibfield  {title} {\enquote {\bibinfo {title}
  {{Quintessence: A Review}},}\ }\href {\doibase
  10.1088/0264-9381/30/21/214003} {\bibfield  {journal} {\bibinfo  {journal}
  {Class. Quant. Grav.}\ }\textbf {\bibinfo {volume} {30}},\ \bibinfo {pages}
  {214003} (\bibinfo {year} {2013})},\ \Eprint {http://arxiv.org/abs/1304.1961}
  {arXiv:1304.1961 [gr-qc]} \BibitemShut {NoStop}%
\bibitem [{\citenamefont {Amendola}\ \emph {et~al.}(2012)\citenamefont
  {Amendola}, \citenamefont {Leite}, \citenamefont {Martins}, \citenamefont
  {Nunes}, \citenamefont {Pedrosa},\ and\ \citenamefont
  {Seganti}}]{Amendola:2011qp}%
  \BibitemOpen
  \bibfield  {author} {\bibinfo {author} {\bibfnamefont {L.}~\bibnamefont
  {Amendola}}, \bibinfo {author} {\bibfnamefont {A.~C.~O.}\ \bibnamefont
  {Leite}}, \bibinfo {author} {\bibfnamefont {C.~J. A.~P.}\ \bibnamefont
  {Martins}}, \bibinfo {author} {\bibfnamefont {N.~J.}\ \bibnamefont {Nunes}},
  \bibinfo {author} {\bibfnamefont {P.~O.~J.}\ \bibnamefont {Pedrosa}}, \ and\
  \bibinfo {author} {\bibfnamefont {A.}~\bibnamefont {Seganti}},\ }\bibfield
  {title} {\enquote {\bibinfo {title} {{Variation of fundamental parameters and
  dark energy. A principal component approach}},}\ }\href {\doibase
  10.1103/PhysRevD.86.063515} {\bibfield  {journal} {\bibinfo  {journal} {Phys.
  Rev.}\ }\textbf {\bibinfo {volume} {D86}},\ \bibinfo {pages} {063515}
  (\bibinfo {year} {2012})},\ \Eprint {http://arxiv.org/abs/1109.6793}
  {arXiv:1109.6793 [astro-ph.CO]} \BibitemShut {NoStop}%
\bibitem [{\citenamefont {Platis}\ \emph {et~al.}(2014)\citenamefont {Platis},
  \citenamefont {Antoniou},\ and\ \citenamefont
  {Perivolaropoulos}}]{Platis:2014sna}%
  \BibitemOpen
  \bibfield  {author} {\bibinfo {author} {\bibfnamefont {Nikos}\ \bibnamefont
  {Platis}}, \bibinfo {author} {\bibfnamefont {Ioannis}\ \bibnamefont
  {Antoniou}}, \ and\ \bibinfo {author} {\bibfnamefont {Leandros}\ \bibnamefont
  {Perivolaropoulos}},\ }\bibfield  {title} {\enquote {\bibinfo {title}
  {{Dilatonic Topological Defects in 3+1 Dimensions and their Embeddings}},}\
  }\href {\doibase 10.1103/PhysRevD.89.123510} {\bibfield  {journal} {\bibinfo
  {journal} {Phys. Rev.}\ }\textbf {\bibinfo {volume} {D89}},\ \bibinfo {pages}
  {123510} (\bibinfo {year} {2014})},\ \Eprint {http://arxiv.org/abs/1406.0639}
  {arXiv:1406.0639 [hep-th]} \BibitemShut {NoStop}%
\bibitem [{\citenamefont {Perivolaropoulos}(2014)}]{Perivolaropoulos:2014lua}%
  \BibitemOpen
  \bibfield  {author} {\bibinfo {author} {\bibfnamefont {Leandros}\
  \bibnamefont {Perivolaropoulos}},\ }\bibfield  {title} {\enquote {\bibinfo
  {title} {{Large Scale Cosmological Anomalies and Inhomogeneous Dark
  Energy}},}\ }\href {\doibase 10.3390/galaxies2010022} {\bibfield  {journal}
  {\bibinfo  {journal} {Galaxies}\ }\textbf {\bibinfo {volume} {2}},\ \bibinfo
  {pages} {22--61} (\bibinfo {year} {2014})},\ \Eprint
  {http://arxiv.org/abs/1401.5044} {arXiv:1401.5044 [astro-ph.CO]} \BibitemShut
  {NoStop}%
\bibitem [{\citenamefont {Mariano}\ and\ \citenamefont
  {Perivolaropoulos}(2013)}]{Mariano:2012ia}%
  \BibitemOpen
  \bibfield  {author} {\bibinfo {author} {\bibfnamefont {Antonio}\ \bibnamefont
  {Mariano}}\ and\ \bibinfo {author} {\bibfnamefont {Leandros}\ \bibnamefont
  {Perivolaropoulos}},\ }\bibfield  {title} {\enquote {\bibinfo {title} {{CMB
  Maximum temperature asymmetry Axis: Alignment with other cosmic
  asymmetries}},}\ }\href {\doibase 10.1103/PhysRevD.87.043511} {\bibfield
  {journal} {\bibinfo  {journal} {Phys. Rev.}\ }\textbf {\bibinfo {volume}
  {D87}},\ \bibinfo {pages} {043511} (\bibinfo {year} {2013})},\ \Eprint
  {http://arxiv.org/abs/1211.5915} {arXiv:1211.5915 [astro-ph.CO]} \BibitemShut
  {NoStop}%
\bibitem [{\citenamefont {King}\ \emph {et~al.}(2012)\citenamefont {King},
  \citenamefont {Webb}, \citenamefont {Murphy}, \citenamefont {Flambaum},
  \citenamefont {Carswell}, \citenamefont {Bainbridge}, \citenamefont
  {Wilczynska},\ and\ \citenamefont {Koch}}]{King:2012id}%
  \BibitemOpen
  \bibfield  {author} {\bibinfo {author} {\bibfnamefont {Julian~A.}\
  \bibnamefont {King}}, \bibinfo {author} {\bibfnamefont {John~K.}\
  \bibnamefont {Webb}}, \bibinfo {author} {\bibfnamefont {Michael~T.}\
  \bibnamefont {Murphy}}, \bibinfo {author} {\bibfnamefont {Victor~V.}\
  \bibnamefont {Flambaum}}, \bibinfo {author} {\bibfnamefont {Robert~F.}\
  \bibnamefont {Carswell}}, \bibinfo {author} {\bibfnamefont {Matthew~B.}\
  \bibnamefont {Bainbridge}}, \bibinfo {author} {\bibfnamefont {Michael~R.}\
  \bibnamefont {Wilczynska}}, \ and\ \bibinfo {author} {\bibfnamefont
  {F.~Elliot}\ \bibnamefont {Koch}},\ }\bibfield  {title} {\enquote {\bibinfo
  {title} {{Spatial variation in the fine-structure constant -- new results
  from VLT/UVES}},}\ }\href {\doibase 10.1111/j.1365-2966.2012.20852.x}
  {\bibfield  {journal} {\bibinfo  {journal} {Mon. Not. Roy. Astron. Soc.}\
  }\textbf {\bibinfo {volume} {422}},\ \bibinfo {pages} {3370--3413} (\bibinfo
  {year} {2012})},\ \Eprint {http://arxiv.org/abs/1202.4758} {arXiv:1202.4758
  [astro-ph.CO]} \BibitemShut {NoStop}%
\bibitem [{\citenamefont {Raffelt}\ and\ \citenamefont
  {Stodolsky}(1988)}]{Raffelt:1987im}%
  \BibitemOpen
  \bibfield  {author} {\bibinfo {author} {\bibfnamefont {Georg}\ \bibnamefont
  {Raffelt}}\ and\ \bibinfo {author} {\bibfnamefont {Leo}\ \bibnamefont
  {Stodolsky}},\ }\bibfield  {title} {\enquote {\bibinfo {title} {{Mixing of
  the Photon with Low Mass Particles}},}\ }\href {\doibase
  10.1103/PhysRevD.37.1237} {\bibfield  {journal} {\bibinfo  {journal} {Phys.
  Rev.}\ }\textbf {\bibinfo {volume} {D37}},\ \bibinfo {pages} {1237} (\bibinfo
  {year} {1988})}\BibitemShut {NoStop}%
\bibitem [{\citenamefont {Duffy}\ and\ \citenamefont {van
  Bibber}(2009)}]{Duffy:2009ig}%
  \BibitemOpen
  \bibfield  {author} {\bibinfo {author} {\bibfnamefont {Leanne~D.}\
  \bibnamefont {Duffy}}\ and\ \bibinfo {author} {\bibfnamefont {Karl}\
  \bibnamefont {van Bibber}},\ }\bibfield  {title} {\enquote {\bibinfo {title}
  {{Axions as Dark Matter Particles}},}\ }\href {\doibase
  10.1088/1367-2630/11/10/105008} {\bibfield  {journal} {\bibinfo  {journal}
  {New J. Phys.}\ }\textbf {\bibinfo {volume} {11}},\ \bibinfo {pages} {105008}
  (\bibinfo {year} {2009})},\ \Eprint {http://arxiv.org/abs/0904.3346}
  {arXiv:0904.3346 [hep-ph]} \BibitemShut {NoStop}%
\bibitem [{\citenamefont {Arias}\ \emph {et~al.}(2012)\citenamefont {Arias},
  \citenamefont {Cadamuro}, \citenamefont {Goodsell}, \citenamefont {Jaeckel},
  \citenamefont {Redondo},\ and\ \citenamefont {Ringwald}}]{Arias:2012az}%
  \BibitemOpen
  \bibfield  {author} {\bibinfo {author} {\bibfnamefont {Paola}\ \bibnamefont
  {Arias}}, \bibinfo {author} {\bibfnamefont {Davide}\ \bibnamefont
  {Cadamuro}}, \bibinfo {author} {\bibfnamefont {Mark}\ \bibnamefont
  {Goodsell}}, \bibinfo {author} {\bibfnamefont {Joerg}\ \bibnamefont
  {Jaeckel}}, \bibinfo {author} {\bibfnamefont {Javier}\ \bibnamefont
  {Redondo}}, \ and\ \bibinfo {author} {\bibfnamefont {Andreas}\ \bibnamefont
  {Ringwald}},\ }\bibfield  {title} {\enquote {\bibinfo {title} {{WISPy Cold
  Dark Matter}},}\ }\href {\doibase 10.1088/1475-7516/2012/06/013} {\bibfield
  {journal} {\bibinfo  {journal} {JCAP}\ }\textbf {\bibinfo {volume} {1206}},\
  \bibinfo {pages} {013} (\bibinfo {year} {2012})},\ \Eprint
  {http://arxiv.org/abs/1201.5902} {arXiv:1201.5902 [hep-ph]} \BibitemShut
  {NoStop}%
\bibitem [{\citenamefont {Cadamuro}(2012)}]{Cadamuro:2012rm}%
  \BibitemOpen
  \bibfield  {author} {\bibinfo {author} {\bibfnamefont {Davide}\ \bibnamefont
  {Cadamuro}},\ }\emph {\bibinfo {title} {{Cosmological limits on axions and
  axion-like particles}}},\ \href
  {https://inspirehep.net/record/1190323/files/arXiv:1210.3196.pdf} {Ph.D.
  thesis},\ \bibinfo  {school} {Munich U.} (\bibinfo {year} {2012}),\ \Eprint
  {http://arxiv.org/abs/1210.3196} {arXiv:1210.3196 [hep-ph]} \BibitemShut
  {NoStop}%
\bibitem [{\citenamefont {Calabrese}\ \emph {et~al.}(2014)\citenamefont
  {Calabrese}, \citenamefont {Martinelli}, \citenamefont {Pandolfi},
  \citenamefont {Cardone}, \citenamefont {Martins}, \citenamefont {Spiro},\
  and\ \citenamefont {Vielzeuf}}]{Calabrese:2013lga}%
  \BibitemOpen
  \bibfield  {author} {\bibinfo {author} {\bibfnamefont {E.}~\bibnamefont
  {Calabrese}}, \bibinfo {author} {\bibfnamefont {M.}~\bibnamefont
  {Martinelli}}, \bibinfo {author} {\bibfnamefont {S.}~\bibnamefont
  {Pandolfi}}, \bibinfo {author} {\bibfnamefont {V.~F.}\ \bibnamefont
  {Cardone}}, \bibinfo {author} {\bibfnamefont {C.~J. A.~P.}\ \bibnamefont
  {Martins}}, \bibinfo {author} {\bibfnamefont {S.}~\bibnamefont {Spiro}}, \
  and\ \bibinfo {author} {\bibfnamefont {P.~E.}\ \bibnamefont {Vielzeuf}},\
  }\bibfield  {title} {\enquote {\bibinfo {title} {{Dark Energy coupling with
  electromagnetism as seen from future low-medium redshift probes}},}\ }\href
  {\doibase 10.1103/PhysRevD.89.083509} {\bibfield  {journal} {\bibinfo
  {journal} {Phys. Rev.}\ }\textbf {\bibinfo {volume} {D89}},\ \bibinfo {pages}
  {083509} (\bibinfo {year} {2014})},\ \Eprint {http://arxiv.org/abs/1311.5841}
  {arXiv:1311.5841 [astro-ph.CO]} \BibitemShut {NoStop}%
\bibitem [{\citenamefont {Calabrese}\ \emph {et~al.}(2011)\citenamefont
  {Calabrese}, \citenamefont {Menegoni}, \citenamefont {Martins}, \citenamefont
  {Melchiorri},\ and\ \citenamefont {Rocha}}]{Calabrese:2011nf}%
  \BibitemOpen
  \bibfield  {author} {\bibinfo {author} {\bibfnamefont {Erminia}\ \bibnamefont
  {Calabrese}}, \bibinfo {author} {\bibfnamefont {Eloisa}\ \bibnamefont
  {Menegoni}}, \bibinfo {author} {\bibfnamefont {C.~J. A.~P.}\ \bibnamefont
  {Martins}}, \bibinfo {author} {\bibfnamefont {Alessandro}\ \bibnamefont
  {Melchiorri}}, \ and\ \bibinfo {author} {\bibfnamefont {Graca}\ \bibnamefont
  {Rocha}},\ }\bibfield  {title} {\enquote {\bibinfo {title} {{Constraining
  Variations in the Fine Structure Constant in the presence of Early Dark
  Energy}},}\ }\href {\doibase 10.1103/PhysRevD.84.023518} {\bibfield
  {journal} {\bibinfo  {journal} {Phys. Rev.}\ }\textbf {\bibinfo {volume}
  {D84}},\ \bibinfo {pages} {023518} (\bibinfo {year} {2011})},\ \Eprint
  {http://arxiv.org/abs/1104.0760} {arXiv:1104.0760 [astro-ph.CO]} \BibitemShut
  {NoStop}%
\bibitem [{\citenamefont {Chou}\ \emph {et~al.}(2008)\citenamefont {Chou},
  \citenamefont {Wester}, \citenamefont {Baumbaugh}, \citenamefont {Gustafson},
  \citenamefont {Irizarry-Valle}, \citenamefont {Mazur}, \citenamefont
  {Steffen}, \citenamefont {Tomlin}, \citenamefont {Yang},\ and\ \citenamefont
  {Yoo}}]{Chou:2007zzc}%
  \BibitemOpen
  \bibfield  {author} {\bibinfo {author} {\bibfnamefont {Aaron~S..}\
  \bibnamefont {Chou}}, \bibinfo {author} {\bibfnamefont {William~Carl}\
  \bibnamefont {Wester}, \bibfnamefont {III}}, \bibinfo {author} {\bibfnamefont
  {A.}~\bibnamefont {Baumbaugh}}, \bibinfo {author} {\bibfnamefont
  {H.~Richard}\ \bibnamefont {Gustafson}}, \bibinfo {author} {\bibfnamefont
  {Y.}~\bibnamefont {Irizarry-Valle}}, \bibinfo {author} {\bibfnamefont
  {P.~O.}\ \bibnamefont {Mazur}}, \bibinfo {author} {\bibfnamefont {Jason~H.}\
  \bibnamefont {Steffen}}, \bibinfo {author} {\bibfnamefont {R.}~\bibnamefont
  {Tomlin}}, \bibinfo {author} {\bibfnamefont {X.}~\bibnamefont {Yang}}, \ and\
  \bibinfo {author} {\bibfnamefont {J.}~\bibnamefont {Yoo}} (\bibinfo
  {collaboration} {GammeV (T-969)}),\ }\bibfield  {title} {\enquote {\bibinfo
  {title} {{Search for axion-like particles using a variable baseline photon
  regeneration technique}},}\ }\href {\doibase 10.1103/PhysRevLett.100.080402}
  {\bibfield  {journal} {\bibinfo  {journal} {Phys. Rev. Lett.}\ }\textbf
  {\bibinfo {volume} {100}},\ \bibinfo {pages} {080402} (\bibinfo {year}
  {2008})},\ \Eprint {http://arxiv.org/abs/0710.3783} {arXiv:0710.3783
  [hep-ex]} \BibitemShut {NoStop}%
\bibitem [{\citenamefont {Masso}\ and\ \citenamefont
  {Toldra}(1995)}]{Masso:1995tw}%
  \BibitemOpen
  \bibfield  {author} {\bibinfo {author} {\bibfnamefont {Eduard}\ \bibnamefont
  {Masso}}\ and\ \bibinfo {author} {\bibfnamefont {Ramon}\ \bibnamefont
  {Toldra}},\ }\bibfield  {title} {\enquote {\bibinfo {title} {{On a light
  spinless particle coupled to photons}},}\ }\href {\doibase
  10.1103/PhysRevD.52.1755} {\bibfield  {journal} {\bibinfo  {journal} {Phys.
  Rev.}\ }\textbf {\bibinfo {volume} {D52}},\ \bibinfo {pages} {1755--1763}
  (\bibinfo {year} {1995})},\ \Eprint {http://arxiv.org/abs/hep-ph/9503293}
  {arXiv:hep-ph/9503293 [hep-ph]} \BibitemShut {NoStop}%
\bibitem [{\citenamefont {Hertzberg}\ \emph {et~al.}(2008)\citenamefont
  {Hertzberg}, \citenamefont {Tegmark},\ and\ \citenamefont
  {Wilczek}}]{Hertzberg:2008wr}%
  \BibitemOpen
  \bibfield  {author} {\bibinfo {author} {\bibfnamefont {Mark~P}\ \bibnamefont
  {Hertzberg}}, \bibinfo {author} {\bibfnamefont {Max}\ \bibnamefont
  {Tegmark}}, \ and\ \bibinfo {author} {\bibfnamefont {Frank}\ \bibnamefont
  {Wilczek}},\ }\bibfield  {title} {\enquote {\bibinfo {title} {{Axion
  Cosmology and the Energy Scale of Inflation}},}\ }\href {\doibase
  10.1103/PhysRevD.78.083507} {\bibfield  {journal} {\bibinfo  {journal} {Phys.
  Rev.}\ }\textbf {\bibinfo {volume} {D78}},\ \bibinfo {pages} {083507}
  (\bibinfo {year} {2008})},\ \Eprint {http://arxiv.org/abs/0807.1726}
  {arXiv:0807.1726 [astro-ph]} \BibitemShut {NoStop}%
\bibitem [{\citenamefont {Asztalos}\ \emph {et~al.}(2004)\citenamefont
  {Asztalos} \emph {et~al.}}]{Asztalos:2003px}%
  \BibitemOpen
  \bibfield  {author} {\bibinfo {author} {\bibfnamefont {Stephen~J.}\
  \bibnamefont {Asztalos}} \emph {et~al.} (\bibinfo {collaboration} {ADMX}),\
  }\bibfield  {title} {\enquote {\bibinfo {title} {{An Improved RF cavity
  search for halo axions}},}\ }\href {\doibase 10.1103/PhysRevD.69.011101}
  {\bibfield  {journal} {\bibinfo  {journal} {Phys. Rev.}\ }\textbf {\bibinfo
  {volume} {D69}},\ \bibinfo {pages} {011101} (\bibinfo {year} {2004})},\
  \Eprint {http://arxiv.org/abs/astro-ph/0310042} {arXiv:astro-ph/0310042
  [astro-ph]} \BibitemShut {NoStop}%
\bibitem [{\citenamefont {Zioutas}\ \emph {et~al.}(2005)\citenamefont {Zioutas}
  \emph {et~al.}}]{Zioutas:2004hi}%
  \BibitemOpen
  \bibfield  {author} {\bibinfo {author} {\bibfnamefont {K.}~\bibnamefont
  {Zioutas}} \emph {et~al.} (\bibinfo {collaboration} {CAST}),\ }\bibfield
  {title} {\enquote {\bibinfo {title} {{First results from the CERN Axion Solar
  Telescope (CAST)}},}\ }\href {\doibase 10.1103/PhysRevLett.94.121301}
  {\bibfield  {journal} {\bibinfo  {journal} {Phys. Rev. Lett.}\ }\textbf
  {\bibinfo {volume} {94}},\ \bibinfo {pages} {121301} (\bibinfo {year}
  {2005})},\ \Eprint {http://arxiv.org/abs/hep-ex/0411033}
  {arXiv:hep-ex/0411033 [hep-ex]} \BibitemShut {NoStop}%
\bibitem [{\citenamefont {Zavattini}\ \emph {et~al.}(2006)\citenamefont
  {Zavattini} \emph {et~al.}}]{Zavattini:2005tm}%
  \BibitemOpen
  \bibfield  {author} {\bibinfo {author} {\bibfnamefont {E.}~\bibnamefont
  {Zavattini}} \emph {et~al.} (\bibinfo {collaboration} {PVLAS}),\ }\bibfield
  {title} {\enquote {\bibinfo {title} {{Experimental observation of optical
  rotation generated in vacuum by a magnetic field}},}\ }\href {\doibase
  10.1103/PhysRevLett.99.129901, 10.1103/PhysRevLett.96.110406} {\bibfield
  {journal} {\bibinfo  {journal} {Phys. Rev. Lett.}\ }\textbf {\bibinfo
  {volume} {96}},\ \bibinfo {pages} {110406} (\bibinfo {year} {2006})},\
  \bibinfo {note} {[Erratum: Phys. Rev. Lett.99,129901(2007)]},\ \Eprint
  {http://arxiv.org/abs/hep-ex/0507107} {arXiv:hep-ex/0507107 [hep-ex]}
  \BibitemShut {NoStop}%
\bibitem [{\citenamefont {Redondo}\ and\ \citenamefont
  {Ringwald}(2011)}]{Redondo:2010dp}%
  \BibitemOpen
  \bibfield  {author} {\bibinfo {author} {\bibfnamefont {Javier}\ \bibnamefont
  {Redondo}}\ and\ \bibinfo {author} {\bibfnamefont {Andreas}\ \bibnamefont
  {Ringwald}},\ }\bibfield  {title} {\enquote {\bibinfo {title} {{Light shining
  through walls}},}\ }\href {\doibase 10.1080/00107514.2011.563516} {\bibfield
  {journal} {\bibinfo  {journal} {Contemp. Phys.}\ }\textbf {\bibinfo {volume}
  {52}},\ \bibinfo {pages} {211--236} (\bibinfo {year} {2011})},\ \Eprint
  {http://arxiv.org/abs/1011.3741} {arXiv:1011.3741 [hep-ph]} \BibitemShut
  {NoStop}%
\bibitem [{\citenamefont {Ruoso}\ \emph {et~al.}(1992)\citenamefont {Ruoso}
  \emph {et~al.}}]{Ruoso:1992nx}%
  \BibitemOpen
  \bibfield  {author} {\bibinfo {author} {\bibfnamefont {G.}~\bibnamefont
  {Ruoso}} \emph {et~al.},\ }\bibfield  {title} {\enquote {\bibinfo {title}
  {{Limits on light scalar and pseudoscalar particles from a photon
  regeneration experiment}},}\ }\href {\doibase 10.1007/BF01474722} {\bibfield
  {journal} {\bibinfo  {journal} {Z. Phys.}\ }\textbf {\bibinfo {volume}
  {C56}},\ \bibinfo {pages} {505--508} (\bibinfo {year} {1992})}\BibitemShut
  {NoStop}%
\bibitem [{\citenamefont {Pugnat}\ \emph {et~al.}(2008)\citenamefont {Pugnat}
  \emph {et~al.}}]{Pugnat:2007nu}%
  \BibitemOpen
  \bibfield  {author} {\bibinfo {author} {\bibfnamefont {Pierre}\ \bibnamefont
  {Pugnat}} \emph {et~al.} (\bibinfo {collaboration} {OSQAR}),\ }\bibfield
  {title} {\enquote {\bibinfo {title} {{First results from the OSQAR photon
  regeneration experiment: No light shining through a wall}},}\ }\href
  {\doibase 10.1103/PhysRevD.78.092003} {\bibfield  {journal} {\bibinfo
  {journal} {Phys. Rev.}\ }\textbf {\bibinfo {volume} {D78}},\ \bibinfo {pages}
  {092003} (\bibinfo {year} {2008})},\ \Eprint {http://arxiv.org/abs/0712.3362}
  {arXiv:0712.3362 [hep-ex]} \BibitemShut {NoStop}%
\bibitem [{\citenamefont {Januschek}(2014)}]{Januschek:2014tua}%
  \BibitemOpen
  \bibfield  {author} {\bibinfo {author} {\bibfnamefont {Friederike}\
  \bibnamefont {Januschek}},\ }\bibfield  {title} {\enquote {\bibinfo {title}
  {{Light-shining-through-walls with lasers}},}\ }in\ \href
  {https://inspirehep.net/record/1320767/files/arXiv:1410.1633.pdf} {\emph
  {\bibinfo {booktitle} {{10th Patras Workshop on Axions, WIMPs and WISPs
  (AXION-WIMP 2014) Geneva, Switzerland, June 29-July 4, 2014}}}}\ (\bibinfo
  {year} {2014})\ \Eprint {http://arxiv.org/abs/1410.1633} {arXiv:1410.1633
  [physics.ins-det]} \BibitemShut {NoStop}%
\bibitem [{\citenamefont {Raffelt}(1996)}]{Raffelt:1996wa}%
  \BibitemOpen
  \bibfield  {author} {\bibinfo {author} {\bibfnamefont {G.~G.}\ \bibnamefont
  {Raffelt}},\ }\href
  {http://wwwth.mpp.mpg.de/members/raffelt/mypapers/199613.pdf} {\emph
  {\bibinfo {title} {{Stars as laboratories for fundamental physics}}}}\
  (\bibinfo {year} {1996})\BibitemShut {NoStop}%
\bibitem [{\citenamefont {Born}\ \emph {et~al.}(1999)\citenamefont {Born},
  \citenamefont {Wolf},\ and\ \citenamefont {Bhatia}}]{born1999principles}%
  \BibitemOpen
  \bibfield  {author} {\bibinfo {author} {\bibfnamefont {M.}~\bibnamefont
  {Born}}, \bibinfo {author} {\bibfnamefont {E.}~\bibnamefont {Wolf}}, \ and\
  \bibinfo {author} {\bibfnamefont {A.B.}\ \bibnamefont {Bhatia}},\ }\href
  {https://books.google.es/books?id=aoX0gYLuENoC} {\emph {\bibinfo {title}
  {Principles of Optics: Electromagnetic Theory of Propagation, Interference
  and Diffraction of Light}}}\ (\bibinfo  {publisher} {Cambridge University
  Press},\ \bibinfo {year} {1999})\BibitemShut {NoStop}%
\bibitem [{\citenamefont {Garcia~de Andrade}(2001)}]{GarciadeAndrade:2001ii}%
  \BibitemOpen
  \bibfield  {author} {\bibinfo {author} {\bibfnamefont {L.~C.}\ \bibnamefont
  {Garcia~de Andrade}},\ }\bibfield  {title} {\enquote {\bibinfo {title}
  {{Cosmic rotation axis, birefrigence and axions to detect primordial torsion
  fields}},}\ }\href@noop {} {\  (\bibinfo {year} {2001})},\ \Eprint
  {http://arxiv.org/abs/hep-th/0110150} {arXiv:hep-th/0110150 [hep-th]}
  \BibitemShut {NoStop}%
\bibitem [{\citenamefont {Villalba-Chávez}(2014)}]{Villalba-Chavez:2013goa}%
  \BibitemOpen
  \bibfield  {author} {\bibinfo {author} {\bibfnamefont {S.}~\bibnamefont
  {Villalba-Chávez}},\ }\bibfield  {title} {\enquote {\bibinfo {title}
  {{Laser-driven search of axion-like particles including vacuum polarization
  effects}},}\ }\href {\doibase 10.1016/j.nuclphysb.2014.01.021} {\bibfield
  {journal} {\bibinfo  {journal} {Nucl. Phys.}\ }\textbf {\bibinfo {volume}
  {B881}},\ \bibinfo {pages} {391--413} (\bibinfo {year} {2014})},\ \Eprint
  {http://arxiv.org/abs/1308.4033} {arXiv:1308.4033 [hep-ph]} \BibitemShut
  {NoStop}%
\bibitem [{\citenamefont {Rubbia}\ and\ \citenamefont
  {Sakharov}(2008)}]{Rubbia:2008us}%
  \BibitemOpen
  \bibfield  {author} {\bibinfo {author} {\bibfnamefont {Andre}\ \bibnamefont
  {Rubbia}}\ and\ \bibinfo {author} {\bibfnamefont {Alexander}\ \bibnamefont
  {Sakharov}},\ }\bibfield  {title} {\enquote {\bibinfo {title} {{Polarization
  mesurements of gamma ray bursts and axion like particles}},}\ }in\ \href
  {\doibase 10.3204/DESY-PROC-2008-02/sakharov_alexander} {\emph {\bibinfo
  {booktitle} {{Axions, WIMPs and WISPs. Proceedings, 4th Patras Workshop,
  PATRAS08, Hamburg, Germany, June 18-21, 2008}}}}\ (\bibinfo {year} {2008})\
  pp.\ \bibinfo {pages} {65--68},\ \Eprint {http://arxiv.org/abs/0809.0612}
  {arXiv:0809.0612 [hep-ph]} \BibitemShut {NoStop}%
\bibitem [{\citenamefont {Antoniadis}\ \emph {et~al.}(2006)\citenamefont
  {Antoniadis}, \citenamefont {Boyarsky},\ and\ \citenamefont
  {Ruchayskiy}}]{Antoniadis:2006wp}%
  \BibitemOpen
  \bibfield  {author} {\bibinfo {author} {\bibfnamefont {I.}~\bibnamefont
  {Antoniadis}}, \bibinfo {author} {\bibfnamefont {Alexey}\ \bibnamefont
  {Boyarsky}}, \ and\ \bibinfo {author} {\bibfnamefont {Oleg}\ \bibnamefont
  {Ruchayskiy}},\ }\bibfield  {title} {\enquote {\bibinfo {title} {{Axion
  alternatives}},}\ }\href@noop {} {\  (\bibinfo {year} {2006})},\ \Eprint
  {http://arxiv.org/abs/hep-ph/0606306} {arXiv:hep-ph/0606306 [hep-ph]}
  \BibitemShut {NoStop}%
\bibitem [{\citenamefont {Dinu}\ \emph {et~al.}(2014)\citenamefont {Dinu},
  \citenamefont {Heinzl}, \citenamefont {Ilderton}, \citenamefont {Marklund},\
  and\ \citenamefont {Torgrimsson}}]{Dinu:2014tsa}%
  \BibitemOpen
  \bibfield  {author} {\bibinfo {author} {\bibfnamefont {Victor}\ \bibnamefont
  {Dinu}}, \bibinfo {author} {\bibfnamefont {Tom}\ \bibnamefont {Heinzl}},
  \bibinfo {author} {\bibfnamefont {Anton}\ \bibnamefont {Ilderton}}, \bibinfo
  {author} {\bibfnamefont {Mattias}\ \bibnamefont {Marklund}}, \ and\ \bibinfo
  {author} {\bibfnamefont {Greger}\ \bibnamefont {Torgrimsson}},\ }\bibfield
  {title} {\enquote {\bibinfo {title} {{Photon polarization in light-by-light
  scattering: Finite size effects}},}\ }\href {\doibase
  10.1103/PhysRevD.90.045025} {\bibfield  {journal} {\bibinfo  {journal} {Phys.
  Rev.}\ }\textbf {\bibinfo {volume} {D90}},\ \bibinfo {pages} {045025}
  (\bibinfo {year} {2014})},\ \Eprint {http://arxiv.org/abs/1405.7291}
  {arXiv:1405.7291 [hep-ph]} \BibitemShut {NoStop}%
\bibitem [{\citenamefont {Antoniadis}\ \emph {et~al.}(2008)\citenamefont
  {Antoniadis}, \citenamefont {Boyarsky},\ and\ \citenamefont
  {Ruchayskiy}}]{Antoniadis:2007sp}%
  \BibitemOpen
  \bibfield  {author} {\bibinfo {author} {\bibfnamefont {Ignatios}\
  \bibnamefont {Antoniadis}}, \bibinfo {author} {\bibfnamefont {Alexey}\
  \bibnamefont {Boyarsky}}, \ and\ \bibinfo {author} {\bibfnamefont {Oleg}\
  \bibnamefont {Ruchayskiy}},\ }\bibfield  {title} {\enquote {\bibinfo {title}
  {{Anomaly induced effects in a magnetic field}},}\ }\href {\doibase
  10.1016/j.nuclphysb.2007.10.006} {\bibfield  {journal} {\bibinfo  {journal}
  {Nucl. Phys.}\ }\textbf {\bibinfo {volume} {B793}},\ \bibinfo {pages}
  {246--259} (\bibinfo {year} {2008})},\ \Eprint
  {http://arxiv.org/abs/0708.3001} {arXiv:0708.3001 [hep-ph]} \BibitemShut
  {NoStop}%
\bibitem [{\citenamefont {Ahlers}\ \emph {et~al.}(2007)\citenamefont {Ahlers},
  \citenamefont {Gies}, \citenamefont {Jaeckel},\ and\ \citenamefont
  {Ringwald}}]{Ahlers:2006iz}%
  \BibitemOpen
  \bibfield  {author} {\bibinfo {author} {\bibfnamefont {Markus}\ \bibnamefont
  {Ahlers}}, \bibinfo {author} {\bibfnamefont {Holger}\ \bibnamefont {Gies}},
  \bibinfo {author} {\bibfnamefont {Joerg}\ \bibnamefont {Jaeckel}}, \ and\
  \bibinfo {author} {\bibfnamefont {Andreas}\ \bibnamefont {Ringwald}},\
  }\bibfield  {title} {\enquote {\bibinfo {title} {{On the Particle
  Interpretation of the PVLAS Data: Neutral versus Charged Particles}},}\
  }\href {\doibase 10.1103/PhysRevD.75.035011} {\bibfield  {journal} {\bibinfo
  {journal} {Phys. Rev.}\ }\textbf {\bibinfo {volume} {D75}},\ \bibinfo {pages}
  {035011} (\bibinfo {year} {2007})},\ \Eprint
  {http://arxiv.org/abs/hep-ph/0612098} {arXiv:hep-ph/0612098 [hep-ph]}
  \BibitemShut {NoStop}%
\bibitem [{\citenamefont {Melchiorri}\ \emph {et~al.}(2007)\citenamefont
  {Melchiorri}, \citenamefont {Polosa},\ and\ \citenamefont
  {Strumia}}]{Melchiorri:2007sq}%
  \BibitemOpen
  \bibfield  {author} {\bibinfo {author} {\bibfnamefont {Alessandro}\
  \bibnamefont {Melchiorri}}, \bibinfo {author} {\bibfnamefont {Antonello}\
  \bibnamefont {Polosa}}, \ and\ \bibinfo {author} {\bibfnamefont {Alessandro}\
  \bibnamefont {Strumia}},\ }\bibfield  {title} {\enquote {\bibinfo {title}
  {{New bounds on millicharged particles from cosmology}},}\ }\href {\doibase
  10.1016/j.physletb.2007.05.042} {\bibfield  {journal} {\bibinfo  {journal}
  {Phys. Lett.}\ }\textbf {\bibinfo {volume} {B650}},\ \bibinfo {pages}
  {416--420} (\bibinfo {year} {2007})},\ \Eprint
  {http://arxiv.org/abs/hep-ph/0703144} {arXiv:hep-ph/0703144 [hep-ph]}
  \BibitemShut {NoStop}%
\bibitem [{\citenamefont {Davidson}\ \emph {et~al.}(2000)\citenamefont
  {Davidson}, \citenamefont {Hannestad},\ and\ \citenamefont
  {Raffelt}}]{Davidson:2000hf}%
  \BibitemOpen
  \bibfield  {author} {\bibinfo {author} {\bibfnamefont {Sacha}\ \bibnamefont
  {Davidson}}, \bibinfo {author} {\bibfnamefont {Steen}\ \bibnamefont
  {Hannestad}}, \ and\ \bibinfo {author} {\bibfnamefont {Georg}\ \bibnamefont
  {Raffelt}},\ }\bibfield  {title} {\enquote {\bibinfo {title} {{Updated bounds
  on millicharged particles}},}\ }\href {\doibase
  10.1088/1126-6708/2000/05/003} {\bibfield  {journal} {\bibinfo  {journal}
  {JHEP}\ }\textbf {\bibinfo {volume} {05}},\ \bibinfo {pages} {003} (\bibinfo
  {year} {2000})},\ \Eprint {http://arxiv.org/abs/hep-ph/0001179}
  {arXiv:hep-ph/0001179 [hep-ph]} \BibitemShut {NoStop}%
\bibitem [{\citenamefont {Zavattini}\ \emph {et~al.}(2008)\citenamefont
  {Zavattini} \emph {et~al.}}]{Zavattini:2007ee}%
  \BibitemOpen
  \bibfield  {author} {\bibinfo {author} {\bibfnamefont {E.}~\bibnamefont
  {Zavattini}} \emph {et~al.} (\bibinfo {collaboration} {PVLAS}),\ }\bibfield
  {title} {\enquote {\bibinfo {title} {{New PVLAS results and limits on
  magnetically induced optical rotation and ellipticity in vacuum}},}\ }\href
  {\doibase 10.1103/PhysRevD.77.032006} {\bibfield  {journal} {\bibinfo
  {journal} {Phys. Rev.}\ }\textbf {\bibinfo {volume} {D77}},\ \bibinfo {pages}
  {032006} (\bibinfo {year} {2008})},\ \Eprint {http://arxiv.org/abs/0706.3419}
  {arXiv:0706.3419 [hep-ex]} \BibitemShut {NoStop}%
\bibitem [{\citenamefont {Brax}\ \emph {et~al.}(2012)\citenamefont {Brax},
  \citenamefont {Burrage},\ and\ \citenamefont {Davis}}]{Brax:2012ie}%
  \BibitemOpen
  \bibfield  {author} {\bibinfo {author} {\bibfnamefont {Philippe}\
  \bibnamefont {Brax}}, \bibinfo {author} {\bibfnamefont {Clare}\ \bibnamefont
  {Burrage}}, \ and\ \bibinfo {author} {\bibfnamefont {Anne-Christine}\
  \bibnamefont {Davis}},\ }\bibfield  {title} {\enquote {\bibinfo {title}
  {{Shining Light on Modifications of Gravity}},}\ }\href {\doibase
  10.1088/1475-7516/2012/10/016} {\bibfield  {journal} {\bibinfo  {journal}
  {JCAP}\ }\textbf {\bibinfo {volume} {1210}},\ \bibinfo {pages} {016}
  (\bibinfo {year} {2012})},\ \Eprint {http://arxiv.org/abs/1206.1809}
  {arXiv:1206.1809 [hep-th]} \BibitemShut {NoStop}%
\bibitem [{\citenamefont {Jaeckel}\ and\ \citenamefont
  {Ringwald}(2010)}]{Jaeckel:2010ni}%
  \BibitemOpen
  \bibfield  {author} {\bibinfo {author} {\bibfnamefont {Joerg}\ \bibnamefont
  {Jaeckel}}\ and\ \bibinfo {author} {\bibfnamefont {Andreas}\ \bibnamefont
  {Ringwald}},\ }\bibfield  {title} {\enquote {\bibinfo {title} {{The
  Low-Energy Frontier of Particle Physics}},}\ }\href {\doibase
  10.1146/annurev.nucl.012809.104433} {\bibfield  {journal} {\bibinfo
  {journal} {Ann. Rev. Nucl. Part. Sci.}\ }\textbf {\bibinfo {volume} {60}},\
  \bibinfo {pages} {405--437} (\bibinfo {year} {2010})},\ \Eprint
  {http://arxiv.org/abs/1002.0329} {arXiv:1002.0329 [hep-ph]} \BibitemShut
  {NoStop}%
\bibitem [{\citenamefont {Dupays}\ \emph {et~al.}(2007)\citenamefont {Dupays},
  \citenamefont {Masso}, \citenamefont {Redondo},\ and\ \citenamefont
  {Rizzo}}]{Dupays:2006dp}%
  \BibitemOpen
  \bibfield  {author} {\bibinfo {author} {\bibfnamefont {Arnaud}\ \bibnamefont
  {Dupays}}, \bibinfo {author} {\bibfnamefont {Eduard}\ \bibnamefont {Masso}},
  \bibinfo {author} {\bibfnamefont {Javier}\ \bibnamefont {Redondo}}, \ and\
  \bibinfo {author} {\bibfnamefont {Carlo}\ \bibnamefont {Rizzo}},\ }\bibfield
  {title} {\enquote {\bibinfo {title} {{Light scalars coupled to photons and
  non-newtonian forces}},}\ }\href {\doibase 10.1103/PhysRevLett.98.131802}
  {\bibfield  {journal} {\bibinfo  {journal} {Phys. Rev. Lett.}\ }\textbf
  {\bibinfo {volume} {98}},\ \bibinfo {pages} {131802} (\bibinfo {year}
  {2007})},\ \Eprint {http://arxiv.org/abs/hep-ph/0610286}
  {arXiv:hep-ph/0610286 [hep-ph]} \BibitemShut {NoStop}%
\bibitem [{\citenamefont {Su}\ \emph {et~al.}(1994)\citenamefont {Su},
  \citenamefont {Heckel}, \citenamefont {Adelberger}, \citenamefont {Gundlach},
  \citenamefont {Harris}, \citenamefont {Smith},\ and\ \citenamefont
  {Swanson}}]{Su:1994gu}%
  \BibitemOpen
  \bibfield  {author} {\bibinfo {author} {\bibfnamefont {Y.}~\bibnamefont
  {Su}}, \bibinfo {author} {\bibfnamefont {Blayne~R.}\ \bibnamefont {Heckel}},
  \bibinfo {author} {\bibfnamefont {E.~G.}\ \bibnamefont {Adelberger}},
  \bibinfo {author} {\bibfnamefont {J.~H.}\ \bibnamefont {Gundlach}}, \bibinfo
  {author} {\bibfnamefont {M.}~\bibnamefont {Harris}}, \bibinfo {author}
  {\bibfnamefont {G.~L.}\ \bibnamefont {Smith}}, \ and\ \bibinfo {author}
  {\bibfnamefont {H.~E.}\ \bibnamefont {Swanson}},\ }\bibfield  {title}
  {\enquote {\bibinfo {title} {{New tests of the universality of free fall}},}\
  }\href {\doibase 10.1103/PhysRevD.50.3614} {\bibfield  {journal} {\bibinfo
  {journal} {Phys. Rev.}\ }\textbf {\bibinfo {volume} {D50}},\ \bibinfo {pages}
  {3614--3636} (\bibinfo {year} {1994})}\BibitemShut {NoStop}%
\bibitem [{\citenamefont {Perivolaropoulos}(2008)}]{Perivolaropoulos:2008pg}%
  \BibitemOpen
  \bibfield  {author} {\bibinfo {author} {\bibfnamefont {Leandros}\
  \bibnamefont {Perivolaropoulos}},\ }\bibfield  {title} {\enquote {\bibinfo
  {title} {{Vacuum energy, the cosmological constant, and compact extra
  dimensions: Constraints from Casimir effect experiments}},}\ }\href {\doibase
  10.1103/PhysRevD.77.107301} {\bibfield  {journal} {\bibinfo  {journal} {Phys.
  Rev.}\ }\textbf {\bibinfo {volume} {D77}},\ \bibinfo {pages} {107301}
  (\bibinfo {year} {2008})},\ \Eprint {http://arxiv.org/abs/0802.1531}
  {arXiv:0802.1531 [astro-ph]} \BibitemShut {NoStop}%
\bibitem [{\citenamefont {Bressi}\ \emph {et~al.}(2002)\citenamefont {Bressi},
  \citenamefont {Carugno}, \citenamefont {Onofrio},\ and\ \citenamefont
  {Ruoso}}]{Bressi:2002fr}%
  \BibitemOpen
  \bibfield  {author} {\bibinfo {author} {\bibfnamefont {G.}~\bibnamefont
  {Bressi}}, \bibinfo {author} {\bibfnamefont {G.}~\bibnamefont {Carugno}},
  \bibinfo {author} {\bibfnamefont {R.}~\bibnamefont {Onofrio}}, \ and\
  \bibinfo {author} {\bibfnamefont {G.}~\bibnamefont {Ruoso}},\ }\bibfield
  {title} {\enquote {\bibinfo {title} {{Measurement of the Casimir force
  between parallel metallic surfaces}},}\ }\href {\doibase
  10.1103/PhysRevLett.88.041804} {\bibfield  {journal} {\bibinfo  {journal}
  {Phys. Rev. Lett.}\ }\textbf {\bibinfo {volume} {88}},\ \bibinfo {pages}
  {041804} (\bibinfo {year} {2002})},\ \Eprint
  {http://arxiv.org/abs/quant-ph/0203002} {arXiv:quant-ph/0203002 [quant-ph]}
  \BibitemShut {NoStop}%
\bibitem [{\citenamefont {Hoyle}\ \emph {et~al.}(2004)\citenamefont {Hoyle},
  \citenamefont {Kapner}, \citenamefont {Heckel}, \citenamefont {Adelberger},
  \citenamefont {Gundlach}, \citenamefont {Schmidt},\ and\ \citenamefont
  {Swanson}}]{Hoyle:2004cw}%
  \BibitemOpen
  \bibfield  {author} {\bibinfo {author} {\bibfnamefont {C.~D.}\ \bibnamefont
  {Hoyle}}, \bibinfo {author} {\bibfnamefont {D.~J.}\ \bibnamefont {Kapner}},
  \bibinfo {author} {\bibfnamefont {Blayne~R.}\ \bibnamefont {Heckel}},
  \bibinfo {author} {\bibfnamefont {E.~G.}\ \bibnamefont {Adelberger}},
  \bibinfo {author} {\bibfnamefont {J.~H.}\ \bibnamefont {Gundlach}}, \bibinfo
  {author} {\bibfnamefont {U.}~\bibnamefont {Schmidt}}, \ and\ \bibinfo
  {author} {\bibfnamefont {H.~E.}\ \bibnamefont {Swanson}},\ }\bibfield
  {title} {\enquote {\bibinfo {title} {{Sub-millimeter tests of the
  gravitational inverse-square law}},}\ }\href {\doibase
  10.1103/PhysRevD.70.042004} {\bibfield  {journal} {\bibinfo  {journal} {Phys.
  Rev.}\ }\textbf {\bibinfo {volume} {D70}},\ \bibinfo {pages} {042004}
  (\bibinfo {year} {2004})},\ \Eprint {http://arxiv.org/abs/hep-ph/0405262}
  {arXiv:hep-ph/0405262 [hep-ph]} \BibitemShut {NoStop}%
\bibitem [{\citenamefont {Adelberger}\ \emph {et~al.}(2007)\citenamefont
  {Adelberger}, \citenamefont {Heckel}, \citenamefont {Hoedl}, \citenamefont
  {Hoyle}, \citenamefont {Kapner},\ and\ \citenamefont
  {Upadhye}}]{Adelberger:2006dh}%
  \BibitemOpen
  \bibfield  {author} {\bibinfo {author} {\bibfnamefont {E.~G.}\ \bibnamefont
  {Adelberger}}, \bibinfo {author} {\bibfnamefont {Blayne~R.}\ \bibnamefont
  {Heckel}}, \bibinfo {author} {\bibfnamefont {Seth~A.}\ \bibnamefont {Hoedl}},
  \bibinfo {author} {\bibfnamefont {C.~D.}\ \bibnamefont {Hoyle}}, \bibinfo
  {author} {\bibfnamefont {D.~J.}\ \bibnamefont {Kapner}}, \ and\ \bibinfo
  {author} {\bibfnamefont {A.}~\bibnamefont {Upadhye}},\ }\bibfield  {title}
  {\enquote {\bibinfo {title} {{Particle Physics Implications of a Recent Test
  of the Gravitational Inverse Sqaure Law}},}\ }\href {\doibase
  10.1103/PhysRevLett.98.131104} {\bibfield  {journal} {\bibinfo  {journal}
  {Phys. Rev. Lett.}\ }\textbf {\bibinfo {volume} {98}},\ \bibinfo {pages}
  {131104} (\bibinfo {year} {2007})},\ \Eprint
  {http://arxiv.org/abs/hep-ph/0611223} {arXiv:hep-ph/0611223 [hep-ph]}
  \BibitemShut {NoStop}%
\bibitem [{\citenamefont {Lee}\ \emph {et~al.}(2004)\citenamefont {Lee},
  \citenamefont {Olive},\ and\ \citenamefont {Pospelov}}]{Lee:2004vm}%
  \BibitemOpen
  \bibfield  {author} {\bibinfo {author} {\bibfnamefont {Seokcheon}\
  \bibnamefont {Lee}}, \bibinfo {author} {\bibfnamefont {Keith~A.}\
  \bibnamefont {Olive}}, \ and\ \bibinfo {author} {\bibfnamefont {Maxim}\
  \bibnamefont {Pospelov}},\ }\bibfield  {title} {\enquote {\bibinfo {title}
  {{Quintessence models and the cosmological evolution of alpha}},}\ }\href
  {\doibase 10.1103/PhysRevD.70.083503} {\bibfield  {journal} {\bibinfo
  {journal} {Phys. Rev.}\ }\textbf {\bibinfo {volume} {D70}},\ \bibinfo {pages}
  {083503} (\bibinfo {year} {2004})},\ \Eprint
  {http://arxiv.org/abs/astro-ph/0406039} {arXiv:astro-ph/0406039 [astro-ph]}
  \BibitemShut {NoStop}%
\bibitem [{\citenamefont {Carroll}(1998)}]{Carroll:1998zi}%
  \BibitemOpen
  \bibfield  {author} {\bibinfo {author} {\bibfnamefont {Sean~M.}\ \bibnamefont
  {Carroll}},\ }\bibfield  {title} {\enquote {\bibinfo {title} {{Quintessence
  and the rest of the world}},}\ }\href {\doibase 10.1103/PhysRevLett.81.3067}
  {\bibfield  {journal} {\bibinfo  {journal} {Phys. Rev. Lett.}\ }\textbf
  {\bibinfo {volume} {81}},\ \bibinfo {pages} {3067--3070} (\bibinfo {year}
  {1998})},\ \Eprint {http://arxiv.org/abs/astro-ph/9806099}
  {arXiv:astro-ph/9806099 [astro-ph]} \BibitemShut {NoStop}%
\bibitem [{\citenamefont {Landau}\ and\ \citenamefont
  {Vucetich}(2002)}]{Landau:2000cc}%
  \BibitemOpen
  \bibfield  {author} {\bibinfo {author} {\bibfnamefont {Susana~J.}\
  \bibnamefont {Landau}}\ and\ \bibinfo {author} {\bibfnamefont {Hector}\
  \bibnamefont {Vucetich}},\ }\bibfield  {title} {\enquote {\bibinfo {title}
  {{Testing theories that predict time variation of fundamental constants}},}\
  }\href {\doibase 10.1086/339775} {\bibfield  {journal} {\bibinfo  {journal}
  {Astrophys. J.}\ }\textbf {\bibinfo {volume} {570}},\ \bibinfo {pages}
  {463--469} (\bibinfo {year} {2002})},\ \Eprint
  {http://arxiv.org/abs/astro-ph/0005316} {arXiv:astro-ph/0005316 [astro-ph]}
  \BibitemShut {NoStop}%
\bibitem [{\citenamefont {Davis}\ \emph {et~al.}(2007)\citenamefont {Davis},
  \citenamefont {Harris}, \citenamefont {Gammon}, \citenamefont {Smolyaninov},\
  and\ \citenamefont {Cho}}]{Davis:2007wu}%
  \BibitemOpen
  \bibfield  {author} {\bibinfo {author} {\bibfnamefont {Christopher~C.}\
  \bibnamefont {Davis}}, \bibinfo {author} {\bibfnamefont {Joseph}\
  \bibnamefont {Harris}}, \bibinfo {author} {\bibfnamefont {Robert~W.}\
  \bibnamefont {Gammon}}, \bibinfo {author} {\bibfnamefont {Igor~I.}\
  \bibnamefont {Smolyaninov}}, \ and\ \bibinfo {author} {\bibfnamefont
  {Kyuman}\ \bibnamefont {Cho}},\ }\bibfield  {title} {\enquote {\bibinfo
  {title} {{Experimental Challenges Involved in Searches for Axion-Like
  Particles and Nonlinear Quantum Electrodynamic Effects by Sensitive Optical
  Techniques}},}\ }\href@noop {} {\  (\bibinfo {year} {2007})},\ \Eprint
  {http://arxiv.org/abs/0704.0748} {arXiv:0704.0748 [hep-th]} \BibitemShut
  {NoStop}%
\bibitem [{\citenamefont {Jaeckel}\ \emph {et~al.}(2007)\citenamefont
  {Jaeckel}, \citenamefont {Masso}, \citenamefont {Redondo}, \citenamefont
  {Ringwald},\ and\ \citenamefont {Takahashi}}]{Jaeckel:2006xm}%
  \BibitemOpen
  \bibfield  {author} {\bibinfo {author} {\bibfnamefont {Joerg}\ \bibnamefont
  {Jaeckel}}, \bibinfo {author} {\bibfnamefont {Eduard}\ \bibnamefont {Masso}},
  \bibinfo {author} {\bibfnamefont {Javier}\ \bibnamefont {Redondo}}, \bibinfo
  {author} {\bibfnamefont {Andreas}\ \bibnamefont {Ringwald}}, \ and\ \bibinfo
  {author} {\bibfnamefont {Fuminobu}\ \bibnamefont {Takahashi}},\ }\bibfield
  {title} {\enquote {\bibinfo {title} {{The Need for purely laboratory-based
  axion-like particle searches}},}\ }\href {\doibase
  10.1103/PhysRevD.75.013004} {\bibfield  {journal} {\bibinfo  {journal} {Phys.
  Rev.}\ }\textbf {\bibinfo {volume} {D75}},\ \bibinfo {pages} {013004}
  (\bibinfo {year} {2007})},\ \Eprint {http://arxiv.org/abs/hep-ph/0610203}
  {arXiv:hep-ph/0610203 [hep-ph]} \BibitemShut {NoStop}%
\bibitem [{\citenamefont {Schott}\ \emph {et~al.}(2011)\citenamefont {Schott}
  \emph {et~al.}}]{Schott:2011fm}%
  \BibitemOpen
  \bibfield  {author} {\bibinfo {author} {\bibfnamefont {Matthias}\
  \bibnamefont {Schott}} \emph {et~al.} (\bibinfo {collaboration} {OSQAR}),\
  }\bibfield  {title} {\enquote {\bibinfo {title} {{First Results of the
  Full-Scale OSQAR Photon Regeneration Experiment}},}\ }in\ \href
  {https://inspirehep.net/record/930419/files/arXiv:1110.0774.pdf} {\emph
  {\bibinfo {booktitle} {{Photon 2011: International Conference On The
  Structure And Interactions Of The Photon and 19th International Workshop On
  Photon-Photon Collisions Spa, Belgium, May 22-27, 2011}}}}\ (\bibinfo {year}
  {2011})\ \Eprint {http://arxiv.org/abs/1110.0774} {arXiv:1110.0774 [hep-ex]}
  \BibitemShut {NoStop}%
\bibitem [{\citenamefont {Ballou}\ \emph {et~al.}(2014)\citenamefont {Ballou}
  \emph {et~al.}}]{Ballou:2014myz}%
  \BibitemOpen
  \bibfield  {author} {\bibinfo {author} {\bibfnamefont {Rafik}\ \bibnamefont
  {Ballou}} \emph {et~al.},\ }\bibfield  {title} {\enquote {\bibinfo {title}
  {{Latest Results of the OSQAR Photon Regeneration Experiment for Axion-Like
  Particle Search}},}\ }in\ \href
  {https://inspirehep.net/record/1321363/files/arXiv:1410.2566.pdf} {\emph
  {\bibinfo {booktitle} {{10th Patras Workshop on Axions, WIMPs and WISPs
  (AXION-WIMP 2014) Geneva, Switzerland, June 29-July 4, 2014}}}}\ (\bibinfo
  {year} {2014})\ \Eprint {http://arxiv.org/abs/1410.2566} {arXiv:1410.2566
  [hep-ex]} \BibitemShut {NoStop}%
\bibitem [{\citenamefont {Ehret}\ \emph {et~al.}(2009)\citenamefont {Ehret}
  \emph {et~al.}}]{Ehret:2009sq}%
  \BibitemOpen
  \bibfield  {author} {\bibinfo {author} {\bibfnamefont {Klaus}\ \bibnamefont
  {Ehret}} \emph {et~al.} (\bibinfo {collaboration} {ALPS}),\ }\bibfield
  {title} {\enquote {\bibinfo {title} {{Resonant laser power build-up in ALPS:
  A 'Light-shining-through-walls' experiment}},}\ }\href {\doibase
  10.1016/j.nima.2009.10.102} {\bibfield  {journal} {\bibinfo  {journal} {Nucl.
  Instrum. Meth.}\ }\textbf {\bibinfo {volume} {A612}},\ \bibinfo {pages}
  {83--96} (\bibinfo {year} {2009})},\ \Eprint {http://arxiv.org/abs/0905.4159}
  {arXiv:0905.4159 [physics.ins-det]} \BibitemShut {NoStop}%
\bibitem [{\citenamefont {Ehret}(2010)}]{Ehret:2010ki}%
  \BibitemOpen
  \bibfield  {author} {\bibinfo {author} {\bibfnamefont {Klaus}\ \bibnamefont
  {Ehret}} (\bibinfo {collaboration} {ALPS}),\ }\bibfield  {title} {\enquote
  {\bibinfo {title} {{The ALPS Light Shining Through a Wall Experiment - WISP
  Search in the Laboratory}},}\ }in\ \href
  {https://inspirehep.net/record/859952/files/arXiv:1006.5741.pdf} {\emph
  {\bibinfo {booktitle} {{Proceedings, 45th Rencontres de Moriond on
  Electroweak Interactions and Unified Theories}}}}\ (\bibinfo {year} {2010})\
  \Eprint {http://arxiv.org/abs/1006.5741} {arXiv:1006.5741 [hep-ex]}
  \BibitemShut {NoStop}%
\bibitem [{\citenamefont {Ehret}\ \emph {et~al.}(2010)\citenamefont {Ehret}
  \emph {et~al.}}]{Ehret:2010mh}%
  \BibitemOpen
  \bibfield  {author} {\bibinfo {author} {\bibfnamefont {Klaus}\ \bibnamefont
  {Ehret}} \emph {et~al.},\ }\bibfield  {title} {\enquote {\bibinfo {title}
  {{New ALPS Results on Hidden-Sector Lightweights}},}\ }\href {\doibase
  10.1016/j.physletb.2010.04.066} {\bibfield  {journal} {\bibinfo  {journal}
  {Phys. Lett.}\ }\textbf {\bibinfo {volume} {B689}},\ \bibinfo {pages}
  {149--155} (\bibinfo {year} {2010})},\ \Eprint
  {http://arxiv.org/abs/1004.1313} {arXiv:1004.1313 [hep-ex]} \BibitemShut
  {NoStop}%
\bibitem [{\citenamefont {Afanasev}\ \emph {et~al.}(2008)\citenamefont
  {Afanasev}, \citenamefont {Baker}, \citenamefont {Beard}, \citenamefont
  {Biallas}, \citenamefont {Boyce}, \citenamefont {Minarni}, \citenamefont
  {Ramdon}, \citenamefont {Shinn},\ and\ \citenamefont
  {Slocum}}]{Afanasev:2008jt}%
  \BibitemOpen
  \bibfield  {author} {\bibinfo {author} {\bibfnamefont {A.}~\bibnamefont
  {Afanasev}}, \bibinfo {author} {\bibfnamefont {O.~K.}\ \bibnamefont {Baker}},
  \bibinfo {author} {\bibfnamefont {K.~B.}\ \bibnamefont {Beard}}, \bibinfo
  {author} {\bibfnamefont {G.}~\bibnamefont {Biallas}}, \bibinfo {author}
  {\bibfnamefont {J.}~\bibnamefont {Boyce}}, \bibinfo {author} {\bibfnamefont
  {M.}~\bibnamefont {Minarni}}, \bibinfo {author} {\bibfnamefont
  {R.}~\bibnamefont {Ramdon}}, \bibinfo {author} {\bibfnamefont
  {M.}~\bibnamefont {Shinn}}, \ and\ \bibinfo {author} {\bibfnamefont
  {P.}~\bibnamefont {Slocum}},\ }\bibfield  {title} {\enquote {\bibinfo {title}
  {{New Experimental limit on Optical Photon Coupling to Neutral, Scalar
  Bosons}},}\ }\href {\doibase 10.1103/PhysRevLett.101.120401} {\bibfield
  {journal} {\bibinfo  {journal} {Phys. Rev. Lett.}\ }\textbf {\bibinfo
  {volume} {101}},\ \bibinfo {pages} {120401} (\bibinfo {year} {2008})},\
  \Eprint {http://arxiv.org/abs/0806.2631} {arXiv:0806.2631 [hep-ex]}
  \BibitemShut {NoStop}%
\bibitem [{\citenamefont {Brax}\ \emph {et~al.}(2004)\citenamefont {Brax},
  \citenamefont {van~de Bruck}, \citenamefont {Davis}, \citenamefont {Khoury},\
  and\ \citenamefont {Weltman}}]{Brax:2004qh}%
  \BibitemOpen
  \bibfield  {author} {\bibinfo {author} {\bibfnamefont {Philippe}\
  \bibnamefont {Brax}}, \bibinfo {author} {\bibfnamefont {Carsten}\
  \bibnamefont {van~de Bruck}}, \bibinfo {author} {\bibfnamefont
  {Anne-Christine}\ \bibnamefont {Davis}}, \bibinfo {author} {\bibfnamefont
  {Justin}\ \bibnamefont {Khoury}}, \ and\ \bibinfo {author} {\bibfnamefont
  {Amanda}\ \bibnamefont {Weltman}},\ }\bibfield  {title} {\enquote {\bibinfo
  {title} {{Detecting dark energy in orbit - The Cosmological chameleon}},}\
  }\href {\doibase 10.1103/PhysRevD.70.123518} {\bibfield  {journal} {\bibinfo
  {journal} {Phys. Rev.}\ }\textbf {\bibinfo {volume} {D70}},\ \bibinfo {pages}
  {123518} (\bibinfo {year} {2004})},\ \Eprint
  {http://arxiv.org/abs/astro-ph/0408415} {arXiv:astro-ph/0408415 [astro-ph]}
  \BibitemShut {NoStop}%
\bibitem [{\citenamefont {Nojiri}\ and\ \citenamefont
  {Odintsov}(2004)}]{Nojiri:2003ti}%
  \BibitemOpen
  \bibfield  {author} {\bibinfo {author} {\bibfnamefont {Shin'ichi}\
  \bibnamefont {Nojiri}}\ and\ \bibinfo {author} {\bibfnamefont {Sergei~D.}\
  \bibnamefont {Odintsov}},\ }\bibfield  {title} {\enquote {\bibinfo {title}
  {{Quantum effects and stability of chameleon cosmology}},}\ }\href {\doibase
  10.1142/S0217732304013933} {\bibfield  {journal} {\bibinfo  {journal} {Mod.
  Phys. Lett.}\ }\textbf {\bibinfo {volume} {A19}},\ \bibinfo {pages}
  {1273--1280} (\bibinfo {year} {2004})},\ \Eprint
  {http://arxiv.org/abs/hep-th/0310296} {arXiv:hep-th/0310296 [hep-th]}
  \BibitemShut {NoStop}%
\bibitem [{\citenamefont {Brax}\ and\ \citenamefont
  {Martin}(2007)}]{Brax:2006np}%
  \BibitemOpen
  \bibfield  {author} {\bibinfo {author} {\bibfnamefont {Philippe}\
  \bibnamefont {Brax}}\ and\ \bibinfo {author} {\bibfnamefont {Jerome}\
  \bibnamefont {Martin}},\ }\bibfield  {title} {\enquote {\bibinfo {title}
  {{Moduli Fields as Quintessence and the Chameleon}},}\ }\href {\doibase
  10.1016/j.physletb.2007.02.019} {\bibfield  {journal} {\bibinfo  {journal}
  {Phys. Lett.}\ }\textbf {\bibinfo {volume} {B647}},\ \bibinfo {pages}
  {320--329} (\bibinfo {year} {2007})},\ \Eprint
  {http://arxiv.org/abs/hep-th/0612208} {arXiv:hep-th/0612208 [hep-th]}
  \BibitemShut {NoStop}%
\bibitem [{\citenamefont {Brax}\ \emph
  {et~al.}(2007{\natexlab{a}})\citenamefont {Brax}, \citenamefont {van~de
  Bruck}, \citenamefont {Davis}, \citenamefont {Mota},\ and\ \citenamefont
  {Shaw}}]{Brax:2007vm}%
  \BibitemOpen
  \bibfield  {author} {\bibinfo {author} {\bibfnamefont {Philippe}\
  \bibnamefont {Brax}}, \bibinfo {author} {\bibfnamefont {Carsten}\
  \bibnamefont {van~de Bruck}}, \bibinfo {author} {\bibfnamefont
  {Anne-Christine}\ \bibnamefont {Davis}}, \bibinfo {author} {\bibfnamefont
  {David~Fonseca}\ \bibnamefont {Mota}}, \ and\ \bibinfo {author}
  {\bibfnamefont {Douglas~J.}\ \bibnamefont {Shaw}},\ }\bibfield  {title}
  {\enquote {\bibinfo {title} {{Detecting chameleons through Casimir force
  measurements}},}\ }\href {\doibase 10.1103/PhysRevD.76.124034} {\bibfield
  {journal} {\bibinfo  {journal} {Phys. Rev.}\ }\textbf {\bibinfo {volume}
  {D76}},\ \bibinfo {pages} {124034} (\bibinfo {year} {2007}{\natexlab{a}})},\
  \Eprint {http://arxiv.org/abs/0709.2075} {arXiv:0709.2075 [hep-ph]}
  \BibitemShut {NoStop}%
\bibitem [{\citenamefont {Nelson}\ and\ \citenamefont
  {Sakellariadou}(2010)}]{Nelson:2008uy}%
  \BibitemOpen
  \bibfield  {author} {\bibinfo {author} {\bibfnamefont {William}\ \bibnamefont
  {Nelson}}\ and\ \bibinfo {author} {\bibfnamefont {Mairi}\ \bibnamefont
  {Sakellariadou}},\ }\bibfield  {title} {\enquote {\bibinfo {title}
  {{Cosmology and the Noncommutative approach to the Standard Model}},}\ }\href
  {\doibase 10.1103/PhysRevD.81.085038} {\bibfield  {journal} {\bibinfo
  {journal} {Phys. Rev.}\ }\textbf {\bibinfo {volume} {D81}},\ \bibinfo {pages}
  {085038} (\bibinfo {year} {2010})},\ \Eprint {http://arxiv.org/abs/0812.1657}
  {arXiv:0812.1657 [hep-th]} \BibitemShut {NoStop}%
\bibitem [{\citenamefont {Brax}\ \emph
  {et~al.}(2007{\natexlab{b}})\citenamefont {Brax}, \citenamefont {van~de
  Bruck},\ and\ \citenamefont {Davis}}]{Brax:2007ak}%
  \BibitemOpen
  \bibfield  {author} {\bibinfo {author} {\bibfnamefont {Philippe}\
  \bibnamefont {Brax}}, \bibinfo {author} {\bibfnamefont {Carsten}\
  \bibnamefont {van~de Bruck}}, \ and\ \bibinfo {author} {\bibfnamefont
  {Anne-Christine}\ \bibnamefont {Davis}},\ }\bibfield  {title} {\enquote
  {\bibinfo {title} {{Compatibility of the chameleon-field model with
  fifth-force experiments, cosmology, and PVLAS and CAST results}},}\ }\href
  {\doibase 10.1103/PhysRevLett.99.121103} {\bibfield  {journal} {\bibinfo
  {journal} {Phys. Rev. Lett.}\ }\textbf {\bibinfo {volume} {99}},\ \bibinfo
  {pages} {121103} (\bibinfo {year} {2007}{\natexlab{b}})},\ \Eprint
  {http://arxiv.org/abs/hep-ph/0703243} {arXiv:hep-ph/0703243 [HEP-PH]}
  \BibitemShut {NoStop}%
\bibitem [{\citenamefont {Hinterbichler}\ \emph {et~al.}(2011)\citenamefont
  {Hinterbichler}, \citenamefont {Khoury},\ and\ \citenamefont
  {Nastase}}]{Hinterbichler:2010wu}%
  \BibitemOpen
  \bibfield  {author} {\bibinfo {author} {\bibfnamefont {Kurt}\ \bibnamefont
  {Hinterbichler}}, \bibinfo {author} {\bibfnamefont {Justin}\ \bibnamefont
  {Khoury}}, \ and\ \bibinfo {author} {\bibfnamefont {Horatiu}\ \bibnamefont
  {Nastase}},\ }\bibfield  {title} {\enquote {\bibinfo {title} {{Towards a UV
  Completion for Chameleon Scalar Theories}},}\ }\href {\doibase
  10.1007/JHEP06(2011)072, 10.1007/JHEP03(2011)061} {\bibfield  {journal}
  {\bibinfo  {journal} {JHEP}\ }\textbf {\bibinfo {volume} {03}},\ \bibinfo
  {pages} {061} (\bibinfo {year} {2011})},\ \bibinfo {note} {[Erratum:
  JHEP06,072(2011)]},\ \Eprint {http://arxiv.org/abs/1012.4462}
  {arXiv:1012.4462 [hep-th]} \BibitemShut {NoStop}%
\bibitem [{\citenamefont {Creminelli}\ \emph {et~al.}(2014)\citenamefont
  {Creminelli}, \citenamefont {Gleyzes}, \citenamefont {Hui}, \citenamefont
  {Simonović},\ and\ \citenamefont {Vernizzi}}]{Creminelli:2013nua}%
  \BibitemOpen
  \bibfield  {author} {\bibinfo {author} {\bibfnamefont {Paolo}\ \bibnamefont
  {Creminelli}}, \bibinfo {author} {\bibfnamefont {Jérôme}\ \bibnamefont
  {Gleyzes}}, \bibinfo {author} {\bibfnamefont {Lam}\ \bibnamefont {Hui}},
  \bibinfo {author} {\bibfnamefont {Marko}\ \bibnamefont {Simonović}}, \ and\
  \bibinfo {author} {\bibfnamefont {Filippo}\ \bibnamefont {Vernizzi}},\
  }\bibfield  {title} {\enquote {\bibinfo {title} {{Single-Field Consistency
  Relations of Large Scale Structure. Part III: Test of the Equivalence
  Principle}},}\ }\href {\doibase 10.1088/1475-7516/2014/06/009} {\bibfield
  {journal} {\bibinfo  {journal} {JCAP}\ }\textbf {\bibinfo {volume} {1406}},\
  \bibinfo {pages} {009} (\bibinfo {year} {2014})},\ \Eprint
  {http://arxiv.org/abs/1312.6074} {arXiv:1312.6074 [astro-ph.CO]} \BibitemShut
  {NoStop}%
\bibitem [{\citenamefont {Joyce}\ \emph {et~al.}(2015)\citenamefont {Joyce},
  \citenamefont {Jain}, \citenamefont {Khoury},\ and\ \citenamefont
  {Trodden}}]{Joyce:2014kja}%
  \BibitemOpen
  \bibfield  {author} {\bibinfo {author} {\bibfnamefont {Austin}\ \bibnamefont
  {Joyce}}, \bibinfo {author} {\bibfnamefont {Bhuvnesh}\ \bibnamefont {Jain}},
  \bibinfo {author} {\bibfnamefont {Justin}\ \bibnamefont {Khoury}}, \ and\
  \bibinfo {author} {\bibfnamefont {Mark}\ \bibnamefont {Trodden}},\ }\bibfield
   {title} {\enquote {\bibinfo {title} {{Beyond the Cosmological Standard
  Model}},}\ }\href {\doibase 10.1016/j.physrep.2014.12.002} {\bibfield
  {journal} {\bibinfo  {journal} {Phys. Rept.}\ }\textbf {\bibinfo {volume}
  {568}},\ \bibinfo {pages} {1--98} (\bibinfo {year} {2015})},\ \Eprint
  {http://arxiv.org/abs/1407.0059} {arXiv:1407.0059 [astro-ph.CO]} \BibitemShut
  {NoStop}%
\bibitem [{\citenamefont {Schelpe}(2010)}]{Schelpe:2010he}%
  \BibitemOpen
  \bibfield  {author} {\bibinfo {author} {\bibfnamefont {Camilla A.~O.}\
  \bibnamefont {Schelpe}},\ }\bibfield  {title} {\enquote {\bibinfo {title}
  {{Chameleon-Photon Mixing in a Primordial Magnetic Field}},}\ }\href
  {\doibase 10.1103/PhysRevD.82.044033} {\bibfield  {journal} {\bibinfo
  {journal} {Phys. Rev.}\ }\textbf {\bibinfo {volume} {D82}},\ \bibinfo {pages}
  {044033} (\bibinfo {year} {2010})},\ \Eprint {http://arxiv.org/abs/1003.0232}
  {arXiv:1003.0232 [astro-ph.CO]} \BibitemShut {NoStop}%
\bibitem [{\citenamefont {Khoury}\ and\ \citenamefont
  {Weltman}(2004{\natexlab{a}})}]{Khoury:2003aq}%
  \BibitemOpen
  \bibfield  {author} {\bibinfo {author} {\bibfnamefont {Justin}\ \bibnamefont
  {Khoury}}\ and\ \bibinfo {author} {\bibfnamefont {Amanda}\ \bibnamefont
  {Weltman}},\ }\bibfield  {title} {\enquote {\bibinfo {title} {{Chameleon
  fields: Awaiting surprises for tests of gravity in space}},}\ }\href
  {\doibase 10.1103/PhysRevLett.93.171104} {\bibfield  {journal} {\bibinfo
  {journal} {Phys. Rev. Lett.}\ }\textbf {\bibinfo {volume} {93}},\ \bibinfo
  {pages} {171104} (\bibinfo {year} {2004}{\natexlab{a}})},\ \Eprint
  {http://arxiv.org/abs/astro-ph/0309300} {arXiv:astro-ph/0309300 [astro-ph]}
  \BibitemShut {NoStop}%
\bibitem [{\citenamefont {Khoury}\ and\ \citenamefont
  {Weltman}(2004{\natexlab{b}})}]{Khoury:2003rn}%
  \BibitemOpen
  \bibfield  {author} {\bibinfo {author} {\bibfnamefont {Justin}\ \bibnamefont
  {Khoury}}\ and\ \bibinfo {author} {\bibfnamefont {Amanda}\ \bibnamefont
  {Weltman}},\ }\bibfield  {title} {\enquote {\bibinfo {title} {{Chameleon
  cosmology}},}\ }\href {\doibase 10.1103/PhysRevD.69.044026} {\bibfield
  {journal} {\bibinfo  {journal} {Phys. Rev.}\ }\textbf {\bibinfo {volume}
  {D69}},\ \bibinfo {pages} {044026} (\bibinfo {year} {2004}{\natexlab{b}})},\
  \Eprint {http://arxiv.org/abs/astro-ph/0309411} {arXiv:astro-ph/0309411
  [astro-ph]} \BibitemShut {NoStop}%
\bibitem [{\citenamefont {Chou}\ \emph {et~al.}(2009)\citenamefont {Chou} \emph
  {et~al.}}]{Chou:2008gr}%
  \BibitemOpen
  \bibfield  {author} {\bibinfo {author} {\bibfnamefont {Aaron~S.}\
  \bibnamefont {Chou}} \emph {et~al.} (\bibinfo {collaboration} {GammeV}),\
  }\bibfield  {title} {\enquote {\bibinfo {title} {{A Search for chameleon
  particles using a photon regeneration technique}},}\ }\href {\doibase
  10.1103/PhysRevLett.102.030402} {\bibfield  {journal} {\bibinfo  {journal}
  {Phys. Rev. Lett.}\ }\textbf {\bibinfo {volume} {102}},\ \bibinfo {pages}
  {030402} (\bibinfo {year} {2009})},\ \Eprint {http://arxiv.org/abs/0806.2438}
  {arXiv:0806.2438 [hep-ex]} \BibitemShut {NoStop}%
\bibitem [{\citenamefont {Ahlers}\ \emph {et~al.}(2008)\citenamefont {Ahlers},
  \citenamefont {Lindner}, \citenamefont {Ringwald}, \citenamefont {Schrempp},\
  and\ \citenamefont {Weniger}}]{Ahlers:2007st}%
  \BibitemOpen
  \bibfield  {author} {\bibinfo {author} {\bibfnamefont {M.}~\bibnamefont
  {Ahlers}}, \bibinfo {author} {\bibfnamefont {A.}~\bibnamefont {Lindner}},
  \bibinfo {author} {\bibfnamefont {A.}~\bibnamefont {Ringwald}}, \bibinfo
  {author} {\bibfnamefont {L.}~\bibnamefont {Schrempp}}, \ and\ \bibinfo
  {author} {\bibfnamefont {C.}~\bibnamefont {Weniger}},\ }\bibfield  {title}
  {\enquote {\bibinfo {title} {{Alpenglow - A Signature for Chameleons in
  Axion-Like Particle Search Experiments}},}\ }\href {\doibase
  10.1103/PhysRevD.77.015018} {\bibfield  {journal} {\bibinfo  {journal} {Phys.
  Rev.}\ }\textbf {\bibinfo {volume} {D77}},\ \bibinfo {pages} {015018}
  (\bibinfo {year} {2008})},\ \Eprint {http://arxiv.org/abs/0710.1555}
  {arXiv:0710.1555 [hep-ph]} \BibitemShut {NoStop}%
\bibitem [{\citenamefont {Steffen}(2008)}]{Steffen:2008au}%
  \BibitemOpen
  \bibfield  {author} {\bibinfo {author} {\bibfnamefont {Jason~H.}\
  \bibnamefont {Steffen}} (\bibinfo {collaboration} {GammeV}),\ }\bibfield
  {title} {\enquote {\bibinfo {title} {{Constraints on chameleons and
  axion-like particles from the GammeV experiment}},}\ }\bibfield  {booktitle}
  {\emph {\bibinfo {booktitle} {{Proceedings, 7th International Workshop on the
  Identification of Dark Matter (IDM 2008)}}},\ }\href@noop {} {\bibfield
  {journal} {\bibinfo  {journal} {PoS}\ }\textbf {\bibinfo {volume}
  {IDM2008}},\ \bibinfo {pages} {064} (\bibinfo {year} {2008})},\ \Eprint
  {http://arxiv.org/abs/0810.5070} {arXiv:0810.5070 [hep-ex]} \BibitemShut
  {NoStop}%
\bibitem [{\citenamefont {Brax}\ \emph {et~al.}(2009)\citenamefont {Brax},
  \citenamefont {van~de Bruck}, \citenamefont {Davis},\ and\ \citenamefont
  {Shaw}}]{Brax:2009bk}%
  \BibitemOpen
  \bibfield  {author} {\bibinfo {author} {\bibfnamefont {Philippe}\
  \bibnamefont {Brax}}, \bibinfo {author} {\bibfnamefont {Carsten}\
  \bibnamefont {van~de Bruck}}, \bibinfo {author} {\bibfnamefont
  {Anne-Christine}\ \bibnamefont {Davis}}, \ and\ \bibinfo {author}
  {\bibfnamefont {Douglas}\ \bibnamefont {Shaw}},\ }\bibfield  {title}
  {\enquote {\bibinfo {title} {{Laboratory Tests of Chameleon Models}},}\ }in\
  \href {\doibase 10.3204/DESY-PROC-2009-05/brax_philippe} {\emph {\bibinfo
  {booktitle} {{Proceedings, 5th Patras Workshop on Axions, WIMPs and WISPs
  (AXION-WIMP 2009)}}}}\ (\bibinfo {year} {2009})\ pp.\ \bibinfo {pages}
  {151--154},\ \Eprint {http://arxiv.org/abs/0911.1086} {arXiv:0911.1086
  [hep-ph]} \BibitemShut {NoStop}%
\bibitem [{\citenamefont {Rybka}\ \emph {et~al.}(2010)\citenamefont {Rybka}
  \emph {et~al.}}]{Rybka:2010ah}%
  \BibitemOpen
  \bibfield  {author} {\bibinfo {author} {\bibfnamefont {G.}~\bibnamefont
  {Rybka}} \emph {et~al.} (\bibinfo {collaboration} {ADMX}),\ }\bibfield
  {title} {\enquote {\bibinfo {title} {{A Search for Scalar Chameleons with
  ADMX}},}\ }\href {\doibase 10.1103/PhysRevLett.105.051801} {\bibfield
  {journal} {\bibinfo  {journal} {Phys. Rev. Lett.}\ }\textbf {\bibinfo
  {volume} {105}},\ \bibinfo {pages} {051801} (\bibinfo {year} {2010})},\
  \Eprint {http://arxiv.org/abs/1004.5160} {arXiv:1004.5160 [astro-ph.CO]}
  \BibitemShut {NoStop}%
\bibitem [{\citenamefont {Steffen}(2010)}]{Steffen:2010ep}%
  \BibitemOpen
  \bibfield  {author} {\bibinfo {author} {\bibfnamefont {Jason~H.}\
  \bibnamefont {Steffen}} (\bibinfo {collaboration} {GammeV-CHASE}),\
  }\bibfield  {title} {\enquote {\bibinfo {title} {{The CHASE laboratory search
  for chameleon dark energy}},}\ }\bibfield  {booktitle} {\emph {\bibinfo
  {booktitle} {{Proceedings, 35th International Conference on High energy
  physics (ICHEP 2010)}}},\ }\href@noop {} {\bibfield  {journal} {\bibinfo
  {journal} {PoS}\ }\textbf {\bibinfo {volume} {ICHEP2010}},\ \bibinfo {pages}
  {446} (\bibinfo {year} {2010})},\ \Eprint {http://arxiv.org/abs/1011.3802}
  {arXiv:1011.3802 [hep-ex]} \BibitemShut {NoStop}%
\bibitem [{\citenamefont {Upadhye}\ \emph {et~al.}(2010)\citenamefont
  {Upadhye}, \citenamefont {Steffen},\ and\ \citenamefont
  {Weltman}}]{Upadhye:2009iv}%
  \BibitemOpen
  \bibfield  {author} {\bibinfo {author} {\bibfnamefont {A.}~\bibnamefont
  {Upadhye}}, \bibinfo {author} {\bibfnamefont {J.~H.}\ \bibnamefont
  {Steffen}}, \ and\ \bibinfo {author} {\bibfnamefont {A.}~\bibnamefont
  {Weltman}},\ }\bibfield  {title} {\enquote {\bibinfo {title} {{Constraining
  chameleon field theories using the GammeV afterglow experiments}},}\ }\href
  {\doibase 10.1103/PhysRevD.81.015013} {\bibfield  {journal} {\bibinfo
  {journal} {Phys. Rev.}\ }\textbf {\bibinfo {volume} {D81}},\ \bibinfo {pages}
  {015013} (\bibinfo {year} {2010})},\ \Eprint {http://arxiv.org/abs/0911.3906}
  {arXiv:0911.3906 [hep-ph]} \BibitemShut {NoStop}%
\bibitem [{\citenamefont {Barrow}\ and\ \citenamefont
  {Graham}(2013)}]{Barrow:2013uza}%
  \BibitemOpen
  \bibfield  {author} {\bibinfo {author} {\bibfnamefont {John~D.}\ \bibnamefont
  {Barrow}}\ and\ \bibinfo {author} {\bibfnamefont {Alexander A.~H.}\
  \bibnamefont {Graham}},\ }\bibfield  {title} {\enquote {\bibinfo {title}
  {{General Dynamics of Varying-Alpha Universes}},}\ }\href {\doibase
  10.1103/PhysRevD.88.103513} {\bibfield  {journal} {\bibinfo  {journal} {Phys.
  Rev.}\ }\textbf {\bibinfo {volume} {D88}},\ \bibinfo {pages} {103513}
  (\bibinfo {year} {2013})},\ \Eprint {http://arxiv.org/abs/1307.6816}
  {arXiv:1307.6816 [gr-qc]} \BibitemShut {NoStop}%
\bibitem [{\citenamefont {Bronnikov}\ \emph {et~al.}(2013)\citenamefont
  {Bronnikov}, \citenamefont {Melnikov}, \citenamefont {Rubin},\ and\
  \citenamefont {Svadkovsky}}]{Bronnikov:2013xh}%
  \BibitemOpen
  \bibfield  {author} {\bibinfo {author} {\bibfnamefont {K.~A.}\ \bibnamefont
  {Bronnikov}}, \bibinfo {author} {\bibfnamefont {V.~N.}\ \bibnamefont
  {Melnikov}}, \bibinfo {author} {\bibfnamefont {S.~G.}\ \bibnamefont {Rubin}},
  \ and\ \bibinfo {author} {\bibfnamefont {I.~V.}\ \bibnamefont {Svadkovsky}},\
  }\bibfield  {title} {\enquote {\bibinfo {title} {{Nonlinear multidimensional
  gravity and the Australian dipole}},}\ }\href {\doibase
  10.1007/s10714-013-1601-2} {\bibfield  {journal} {\bibinfo  {journal} {Gen.
  Rel. Grav.}\ }\textbf {\bibinfo {volume} {45}},\ \bibinfo {pages}
  {2509--2528} (\bibinfo {year} {2013})},\ \Eprint
  {http://arxiv.org/abs/1301.3098} {arXiv:1301.3098 [gr-qc]} \BibitemShut
  {NoStop}%
\bibitem [{\citenamefont {Perivolaropoulos}(2005)}]{Perivolaropoulos:2004yr}%
  \BibitemOpen
  \bibfield  {author} {\bibinfo {author} {\bibfnamefont {Leandros}\
  \bibnamefont {Perivolaropoulos}},\ }\bibfield  {title} {\enquote {\bibinfo
  {title} {{Constraints on linear negative potentials in quintessence and
  phantom models from recent supernova data}},}\ }\href {\doibase
  10.1103/PhysRevD.71.063503} {\bibfield  {journal} {\bibinfo  {journal} {Phys.
  Rev.}\ }\textbf {\bibinfo {volume} {D71}},\ \bibinfo {pages} {063503}
  (\bibinfo {year} {2005})},\ \Eprint {http://arxiv.org/abs/astro-ph/0412308}
  {arXiv:astro-ph/0412308 [astro-ph]} \BibitemShut {NoStop}%
\bibitem [{\citenamefont {Kallosh}\ \emph {et~al.}(2003)\citenamefont
  {Kallosh}, \citenamefont {Kratochvil}, \citenamefont {Linde}, \citenamefont
  {Linder},\ and\ \citenamefont {Shmakova}}]{Kallosh:2003bq}%
  \BibitemOpen
  \bibfield  {author} {\bibinfo {author} {\bibfnamefont {Renata}\ \bibnamefont
  {Kallosh}}, \bibinfo {author} {\bibfnamefont {Jan}\ \bibnamefont
  {Kratochvil}}, \bibinfo {author} {\bibfnamefont {Andrei~D.}\ \bibnamefont
  {Linde}}, \bibinfo {author} {\bibfnamefont {Eric~V.}\ \bibnamefont {Linder}},
  \ and\ \bibinfo {author} {\bibfnamefont {Marina}\ \bibnamefont {Shmakova}},\
  }\bibfield  {title} {\enquote {\bibinfo {title} {{Observational bounds on
  cosmic doomsday}},}\ }\href {\doibase 10.1088/1475-7516/2003/10/015}
  {\bibfield  {journal} {\bibinfo  {journal} {JCAP}\ }\textbf {\bibinfo
  {volume} {0310}},\ \bibinfo {pages} {015} (\bibinfo {year} {2003})},\ \Eprint
  {http://arxiv.org/abs/astro-ph/0307185} {arXiv:astro-ph/0307185 [astro-ph]}
  \BibitemShut {NoStop}%
\bibitem [{\citenamefont {Lykkas}\ and\ \citenamefont
  {Perivolaropoulos}(2016)}]{Lykkas:2015kls}%
  \BibitemOpen
  \bibfield  {author} {\bibinfo {author} {\bibfnamefont {A.}~\bibnamefont
  {Lykkas}}\ and\ \bibinfo {author} {\bibfnamefont {L.}~\bibnamefont
  {Perivolaropoulos}},\ }\bibfield  {title} {\enquote {\bibinfo {title}
  {{Scalar-Tensor Quintessence with a linear potential: Avoiding the Big Crunch
  cosmic doomsday}},}\ }\href {\doibase 10.1103/PhysRevD.93.043513} {\bibfield
  {journal} {\bibinfo  {journal} {Phys. Rev.}\ }\textbf {\bibinfo {volume}
  {D93}},\ \bibinfo {pages} {043513} (\bibinfo {year} {2016})},\ \Eprint
  {http://arxiv.org/abs/1511.08732} {arXiv:1511.08732 [gr-qc]} \BibitemShut
  {NoStop}%
\bibitem [{\citenamefont {Bronnikov}\ \emph {et~al.}(2010)\citenamefont
  {Bronnikov}, \citenamefont {Rubin},\ and\ \citenamefont
  {Svadkovsky}}]{Bronnikov:2009ai}%
  \BibitemOpen
  \bibfield  {author} {\bibinfo {author} {\bibfnamefont {K.~A.}\ \bibnamefont
  {Bronnikov}}, \bibinfo {author} {\bibfnamefont {S.~G.}\ \bibnamefont
  {Rubin}}, \ and\ \bibinfo {author} {\bibfnamefont {I.~V.}\ \bibnamefont
  {Svadkovsky}},\ }\bibfield  {title} {\enquote {\bibinfo {title}
  {{Multidimensional world, inflation and modern acceleration}},}\ }\href
  {\doibase 10.1103/PhysRevD.81.084010} {\bibfield  {journal} {\bibinfo
  {journal} {Phys. Rev.}\ }\textbf {\bibinfo {volume} {D81}},\ \bibinfo {pages}
  {084010} (\bibinfo {year} {2010})},\ \Eprint {http://arxiv.org/abs/0912.4862}
  {arXiv:0912.4862 [gr-qc]} \BibitemShut {NoStop}%
\bibitem [{\citenamefont {Bronnikov}\ and\ \citenamefont
  {Melnikov}(2003)}]{Bronnikov:2003rf}%
  \BibitemOpen
  \bibfield  {author} {\bibinfo {author} {\bibfnamefont {K.~A.}\ \bibnamefont
  {Bronnikov}}\ and\ \bibinfo {author} {\bibfnamefont {V.~N.}\ \bibnamefont
  {Melnikov}},\ }\bibfield  {title} {\enquote {\bibinfo {title} {{Conformal
  frames and D-dimensional gravity}},}\ }in\ \href {\doibase
  10.1007/978-1-4020-2242-5_2} {\emph {\bibinfo {booktitle} {{International
  School of Cosmology and Gravitation: 18th Course: The Gravitational Constant:
  Generalized Gravitational Theories and Experiments: A NATO Advanced Study
  Institute Erice, Italy, April 30-May 10, 2003}}}}\ (\bibinfo {year} {2003})\
  pp.\ \bibinfo {pages} {39--64},\ \Eprint {http://arxiv.org/abs/gr-qc/0310112}
  {arXiv:gr-qc/0310112 [gr-qc]} \BibitemShut {NoStop}%
\bibitem [{\citenamefont {Bronnikov}\ and\ \citenamefont
  {Melnikov}(2001)}]{Bronnikov:2001th}%
  \BibitemOpen
  \bibfield  {author} {\bibinfo {author} {\bibfnamefont {K.~A.}\ \bibnamefont
  {Bronnikov}}\ and\ \bibinfo {author} {\bibfnamefont {V.~N.}\ \bibnamefont
  {Melnikov}},\ }\bibfield  {title} {\enquote {\bibinfo {title} {{On
  observational predictions from multidimensional gravity}},}\ }\href {\doibase
  10.1023/A:1012245011428} {\bibfield  {journal} {\bibinfo  {journal} {Gen.
  Rel. Grav.}\ }\textbf {\bibinfo {volume} {33}},\ \bibinfo {pages}
  {1549--1578} (\bibinfo {year} {2001})},\ \Eprint
  {http://arxiv.org/abs/gr-qc/0103079} {arXiv:gr-qc/0103079 [gr-qc]}
  \BibitemShut {NoStop}%
\bibitem [{\citenamefont {Pettini}\ \emph {et~al.}(2001)\citenamefont
  {Pettini}, \citenamefont {Shapley}, \citenamefont {Steidel}, \citenamefont
  {Cuby}, \citenamefont {Dickinson}, \citenamefont {Moorwood}, \citenamefont
  {Adelberger},\ and\ \citenamefont {Giavalisco}}]{Pettini:2001wp}%
  \BibitemOpen
  \bibfield  {author} {\bibinfo {author} {\bibfnamefont {Max}\ \bibnamefont
  {Pettini}}, \bibinfo {author} {\bibfnamefont {Alice~E.}\ \bibnamefont
  {Shapley}}, \bibinfo {author} {\bibfnamefont {Charles~C.}\ \bibnamefont
  {Steidel}}, \bibinfo {author} {\bibfnamefont {Jean-Gabriel}\ \bibnamefont
  {Cuby}}, \bibinfo {author} {\bibfnamefont {Mark}\ \bibnamefont {Dickinson}},
  \bibinfo {author} {\bibfnamefont {Alan F.~M.}\ \bibnamefont {Moorwood}},
  \bibinfo {author} {\bibfnamefont {Kurt~L.}\ \bibnamefont {Adelberger}}, \
  and\ \bibinfo {author} {\bibfnamefont {Mauro}\ \bibnamefont {Giavalisco}},\
  }\bibfield  {title} {\enquote {\bibinfo {title} {{The Rest frame optical
  spectra of Lyman break galaxies: Star formation, extinction, abundances, and
  kinematics}},}\ }\href {\doibase 10.1086/321403} {\bibfield  {journal}
  {\bibinfo  {journal} {Astrophys. J.}\ }\textbf {\bibinfo {volume} {554}},\
  \bibinfo {pages} {981--1000} (\bibinfo {year} {2001})},\ \Eprint
  {http://arxiv.org/abs/astro-ph/0102456} {arXiv:astro-ph/0102456 [astro-ph]}
  \BibitemShut {NoStop}%
\bibitem [{\citenamefont {Vogt}\ \emph {et~al.}(1994)\citenamefont {Vogt} \emph
  {et~al.}}]{Vogt:1995zz}%
  \BibitemOpen
  \bibfield  {author} {\bibinfo {author} {\bibfnamefont {S.~S.}\ \bibnamefont
  {Vogt}} \emph {et~al.},\ }\bibfield  {title} {\enquote {\bibinfo {title}
  {{HIRES: the high-resolution echelle spectrometer on the Keck 10-m
  Telescope}},}\ }\href {\doibase 10.1117/12.176725} {\bibfield  {journal}
  {\bibinfo  {journal} {Proc. SPIE Int. Soc. Opt. Eng.}\ }\textbf {\bibinfo
  {volume} {2198}},\ \bibinfo {pages} {362} (\bibinfo {year}
  {1994})}\BibitemShut {NoStop}%
\bibitem [{\citenamefont {Faber}\ \emph {et~al.}(1969)\citenamefont {Faber}
  \emph {et~al.}}]{Faber:2003zz}%
  \BibitemOpen
  \bibfield  {author} {\bibinfo {author} {\bibfnamefont {Sandra~M.}\
  \bibnamefont {Faber}} \emph {et~al.},\ }\bibfield  {title} {\enquote
  {\bibinfo {title} {{The DEIMOS spectrograph for the Keck II Telescope:
  integration and testing}},}\ }\href {\doibase 10.1117/12.460346} {\bibfield
  {journal} {\bibinfo  {journal} {Proc. SPIE Int. Soc. Opt. Eng.}\ }\textbf
  {\bibinfo {volume} {4841}},\ \bibinfo {pages} {1657--1669} (\bibinfo {year}
  {1969})}\BibitemShut {NoStop}%
\bibitem [{\citenamefont {Weber}\ and\ \citenamefont
  {de~Boer}(2010)}]{Weber:2009pt}%
  \BibitemOpen
  \bibfield  {author} {\bibinfo {author} {\bibfnamefont {Markus}\ \bibnamefont
  {Weber}}\ and\ \bibinfo {author} {\bibfnamefont {Wim}\ \bibnamefont
  {de~Boer}},\ }\bibfield  {title} {\enquote {\bibinfo {title} {{Determination
  of the Local Dark Matter Density in our Galaxy}},}\ }\href {\doibase
  10.1051/0004-6361/200913381} {\bibfield  {journal} {\bibinfo  {journal}
  {Astron. Astrophys.}\ }\textbf {\bibinfo {volume} {509}},\ \bibinfo {pages}
  {A25} (\bibinfo {year} {2010})},\ \Eprint {http://arxiv.org/abs/0910.4272}
  {arXiv:0910.4272 [astro-ph.CO]} \BibitemShut {NoStop}%
\bibitem [{\citenamefont {Asztalos}\ \emph {et~al.}(2002)\citenamefont
  {Asztalos} \emph {et~al.}}]{Asztalos:2001jk}%
  \BibitemOpen
  \bibfield  {author} {\bibinfo {author} {\bibfnamefont {Stephen~J.}\
  \bibnamefont {Asztalos}} \emph {et~al.} (\bibinfo {collaboration} {ADMX}),\
  }\bibfield  {title} {\enquote {\bibinfo {title} {{Experimental constraints on
  the axion dark matter halo density}},}\ }\href {\doibase 10.1086/341130}
  {\bibfield  {journal} {\bibinfo  {journal} {Astrophys. J.}\ }\textbf
  {\bibinfo {volume} {571}},\ \bibinfo {pages} {L27--L30} (\bibinfo {year}
  {2002})},\ \Eprint {http://arxiv.org/abs/astro-ph/0104200}
  {arXiv:astro-ph/0104200 [astro-ph]} \BibitemShut {NoStop}%
\bibitem [{\citenamefont {de~Boer}\ and\ \citenamefont
  {Weber}(2011)}]{deBoer:2010eh}%
  \BibitemOpen
  \bibfield  {author} {\bibinfo {author} {\bibfnamefont {W.}~\bibnamefont
  {de~Boer}}\ and\ \bibinfo {author} {\bibfnamefont {M.}~\bibnamefont
  {Weber}},\ }\bibfield  {title} {\enquote {\bibinfo {title} {{The Dark Matter
  Density in the Solar Neighborhood reconsidered}},}\ }\href {\doibase
  10.1088/1475-7516/2011/04/002} {\bibfield  {journal} {\bibinfo  {journal}
  {JCAP}\ }\textbf {\bibinfo {volume} {1104}},\ \bibinfo {pages} {002}
  (\bibinfo {year} {2011})},\ \Eprint {http://arxiv.org/abs/1011.6323}
  {arXiv:1011.6323 [astro-ph.CO]} \BibitemShut {NoStop}%
\bibitem [{\citenamefont {Marsh}(2015)}]{Marsh:2014gca}%
  \BibitemOpen
  \bibfield  {author} {\bibinfo {author} {\bibfnamefont {M.~C.~David}\
  \bibnamefont {Marsh}},\ }\bibfield  {title} {\enquote {\bibinfo {title} {{The
  Darkness of Spin-0 Dark Radiation}},}\ }\href {\doibase
  10.1088/1475-7516/2015/01/017} {\bibfield  {journal} {\bibinfo  {journal}
  {JCAP}\ }\textbf {\bibinfo {volume} {1501}},\ \bibinfo {pages} {017}
  (\bibinfo {year} {2015})},\ \Eprint {http://arxiv.org/abs/1407.2501}
  {arXiv:1407.2501 [hep-ph]} \BibitemShut {NoStop}%
\bibitem [{\citenamefont {Giannotti}\ \emph {et~al.}(2016)\citenamefont
  {Giannotti}, \citenamefont {Irastorza}, \citenamefont {Redondo},\ and\
  \citenamefont {Ringwald}}]{Giannotti:2015kwo}%
  \BibitemOpen
  \bibfield  {author} {\bibinfo {author} {\bibfnamefont {Maurizio}\
  \bibnamefont {Giannotti}}, \bibinfo {author} {\bibfnamefont {Igor}\
  \bibnamefont {Irastorza}}, \bibinfo {author} {\bibfnamefont {Javier}\
  \bibnamefont {Redondo}}, \ and\ \bibinfo {author} {\bibfnamefont {Andreas}\
  \bibnamefont {Ringwald}},\ }\bibfield  {title} {\enquote {\bibinfo {title}
  {{Cool WISPs for stellar cooling excesses}},}\ }\href {\doibase
  10.1088/1475-7516/2016/05/057} {\bibfield  {journal} {\bibinfo  {journal}
  {JCAP}\ }\textbf {\bibinfo {volume} {1605}},\ \bibinfo {pages} {057}
  (\bibinfo {year} {2016})},\ \Eprint {http://arxiv.org/abs/1512.08108}
  {arXiv:1512.08108 [astro-ph.HE]} \BibitemShut {NoStop}%
\bibitem [{\citenamefont {Beringer}\ \emph {et~al.}(2012)\citenamefont
  {Beringer} \emph {et~al.}}]{Beringer:1900zz}%
  \BibitemOpen
  \bibfield  {author} {\bibinfo {author} {\bibfnamefont {J.}~\bibnamefont
  {Beringer}} \emph {et~al.} (\bibinfo {collaboration} {Particle Data Group}),\
  }\bibfield  {title} {\enquote {\bibinfo {title} {{Review of Particle Physics
  (RPP)}},}\ }\href {\doibase 10.1103/PhysRevD.86.010001} {\bibfield  {journal}
  {\bibinfo  {journal} {Phys. Rev.}\ }\textbf {\bibinfo {volume} {D86}},\
  \bibinfo {pages} {010001} (\bibinfo {year} {2012})}\BibitemShut {NoStop}%
\bibitem [{\citenamefont {Payez}\ \emph {et~al.}(2011)\citenamefont {Payez},
  \citenamefont {Cudell},\ and\ \citenamefont {Hutsemekers}}]{Payez:2011sh}%
  \BibitemOpen
  \bibfield  {author} {\bibinfo {author} {\bibfnamefont {A.}~\bibnamefont
  {Payez}}, \bibinfo {author} {\bibfnamefont {J.~R.}\ \bibnamefont {Cudell}}, \
  and\ \bibinfo {author} {\bibfnamefont {D.}~\bibnamefont {Hutsemekers}},\
  }\bibfield  {title} {\enquote {\bibinfo {title} {{Can axion-like particles
  explain the alignments of the polarisations of light from quasars?}}}\ }\href
  {\doibase 10.1103/PhysRevD.84.085029} {\bibfield  {journal} {\bibinfo
  {journal} {Phys. Rev.}\ }\textbf {\bibinfo {volume} {D84}},\ \bibinfo {pages}
  {085029} (\bibinfo {year} {2011})},\ \Eprint {http://arxiv.org/abs/1107.2013}
  {arXiv:1107.2013 [astro-ph.CO]} \BibitemShut {NoStop}%
\bibitem [{\citenamefont {Zioutas}\ \emph {et~al.}(2006)\citenamefont {Zioutas}
  \emph {et~al.}}]{Zioutas:2006ns}%
  \BibitemOpen
  \bibfield  {author} {\bibinfo {author} {\bibfnamefont {Konstantin}\
  \bibnamefont {Zioutas}} \emph {et~al.},\ }\bibfield  {title} {\enquote
  {\bibinfo {title} {{Indirect signatures for axion(-like) particles}},}\
  }\bibfield  {booktitle} {\emph {\bibinfo {booktitle} {{TAUP 2005.
  Proceedings, 9th International Conference on topics in astroparticle and
  underground physics, Zaragoza, September 10, 2005}}},\ }\href {\doibase
  10.1088/1742-6596/39/1/020} {\bibfield  {journal} {\bibinfo  {journal} {J.
  Phys. Conf. Ser.}\ }\textbf {\bibinfo {volume} {39}},\ \bibinfo {pages}
  {103--106} (\bibinfo {year} {2006})},\ \Eprint
  {http://arxiv.org/abs/astro-ph/0603507} {arXiv:astro-ph/0603507 [astro-ph]}
  \BibitemShut {NoStop}%
\bibitem [{\citenamefont {Gorbunov}\ \emph {et~al.}(2001)\citenamefont
  {Gorbunov}, \citenamefont {Raffelt},\ and\ \citenamefont
  {Semikoz}}]{Gorbunov:2001gc}%
  \BibitemOpen
  \bibfield  {author} {\bibinfo {author} {\bibfnamefont {D.~S.}\ \bibnamefont
  {Gorbunov}}, \bibinfo {author} {\bibfnamefont {G.~G.}\ \bibnamefont
  {Raffelt}}, \ and\ \bibinfo {author} {\bibfnamefont {D.~V.}\ \bibnamefont
  {Semikoz}},\ }\bibfield  {title} {\enquote {\bibinfo {title} {{Axion - like
  particles as ultrahigh-energy cosmic rays?}}}\ }\href {\doibase
  10.1103/PhysRevD.64.096005} {\bibfield  {journal} {\bibinfo  {journal} {Phys.
  Rev.}\ }\textbf {\bibinfo {volume} {D64}},\ \bibinfo {pages} {096005}
  (\bibinfo {year} {2001})},\ \Eprint {http://arxiv.org/abs/hep-ph/0103175}
  {arXiv:hep-ph/0103175 [hep-ph]} \BibitemShut {NoStop}%
\bibitem [{\citenamefont {Burrage}\ \emph
  {et~al.}(2009{\natexlab{a}})\citenamefont {Burrage}, \citenamefont {Davis},\
  and\ \citenamefont {Shaw}}]{Burrage:2009mj}%
  \BibitemOpen
  \bibfield  {author} {\bibinfo {author} {\bibfnamefont {Clare}\ \bibnamefont
  {Burrage}}, \bibinfo {author} {\bibfnamefont {Anne-Christine}\ \bibnamefont
  {Davis}}, \ and\ \bibinfo {author} {\bibfnamefont {Douglas~J.}\ \bibnamefont
  {Shaw}},\ }\bibfield  {title} {\enquote {\bibinfo {title} {{Active Galactic
  Nuclei Shed Light on Axion-like-Particles}},}\ }\href {\doibase
  10.1103/PhysRevLett.102.201101} {\bibfield  {journal} {\bibinfo  {journal}
  {Phys. Rev. Lett.}\ }\textbf {\bibinfo {volume} {102}},\ \bibinfo {pages}
  {201101} (\bibinfo {year} {2009}{\natexlab{a}})},\ \Eprint
  {http://arxiv.org/abs/0902.2320} {arXiv:0902.2320 [astro-ph.CO]} \BibitemShut
  {NoStop}%
\bibitem [{\citenamefont {Payez}\ \emph {et~al.}(2008)\citenamefont {Payez},
  \citenamefont {Cudell},\ and\ \citenamefont {Hutsemekers}}]{Payez:2008pm}%
  \BibitemOpen
  \bibfield  {author} {\bibinfo {author} {\bibfnamefont {A.}~\bibnamefont
  {Payez}}, \bibinfo {author} {\bibfnamefont {J.~R.}\ \bibnamefont {Cudell}}, \
  and\ \bibinfo {author} {\bibfnamefont {D.}~\bibnamefont {Hutsemekers}},\
  }\bibfield  {title} {\enquote {\bibinfo {title} {{Axions and polarisation of
  quasars}},}\ }\bibfield  {booktitle} {\emph {\bibinfo {booktitle} {{Hadronic
  physics. Proceedings, Joint Meeting Heidelberg-Liege-Paris-Wroclaw, HLPW
  2008, Spa, Belgium, March 6-8, 2008}}},\ }\href {\doibase 10.1063/1.2987174}
  {\bibfield  {journal} {\bibinfo  {journal} {AIP Conf. Proc.}\ }\textbf
  {\bibinfo {volume} {1038}},\ \bibinfo {pages} {211--219} (\bibinfo {year}
  {2008})},\ \Eprint {http://arxiv.org/abs/0805.3946} {arXiv:0805.3946
  [astro-ph]} \BibitemShut {NoStop}%
\bibitem [{\citenamefont {Zioutas}\ \emph {et~al.}(2009)\citenamefont
  {Zioutas}, \citenamefont {Tsagri}, \citenamefont {Semertzidis}, \citenamefont
  {Papaevangelou}, \citenamefont {Dafni},\ and\ \citenamefont
  {Anastassopoulos}}]{Zioutas:2009bw}%
  \BibitemOpen
  \bibfield  {author} {\bibinfo {author} {\bibfnamefont {K.}~\bibnamefont
  {Zioutas}}, \bibinfo {author} {\bibfnamefont {M.}~\bibnamefont {Tsagri}},
  \bibinfo {author} {\bibfnamefont {Y.}~\bibnamefont {Semertzidis}}, \bibinfo
  {author} {\bibfnamefont {T.}~\bibnamefont {Papaevangelou}}, \bibinfo {author}
  {\bibfnamefont {T.}~\bibnamefont {Dafni}}, \ and\ \bibinfo {author}
  {\bibfnamefont {V.}~\bibnamefont {Anastassopoulos}},\ }\bibfield  {title}
  {\enquote {\bibinfo {title} {{Axion Searches with Helioscopes and
  astrophysical signatures for axion(-like) particles}},}\ }\href {\doibase
  10.1088/1367-2630/11/10/105020} {\bibfield  {journal} {\bibinfo  {journal}
  {New J. Phys.}\ }\textbf {\bibinfo {volume} {11}},\ \bibinfo {pages} {105020}
  (\bibinfo {year} {2009})},\ \Eprint {http://arxiv.org/abs/0903.1807}
  {arXiv:0903.1807 [astro-ph.SR]} \BibitemShut {NoStop}%
\bibitem [{\citenamefont {Fairbairn}\ \emph {et~al.}(2011)\citenamefont
  {Fairbairn}, \citenamefont {Rashba},\ and\ \citenamefont
  {Troitsky}}]{Fairbairn:2009zi}%
  \BibitemOpen
  \bibfield  {author} {\bibinfo {author} {\bibfnamefont {Malcolm}\ \bibnamefont
  {Fairbairn}}, \bibinfo {author} {\bibfnamefont {Timur}\ \bibnamefont
  {Rashba}}, \ and\ \bibinfo {author} {\bibfnamefont {Sergey~V.}\ \bibnamefont
  {Troitsky}},\ }\bibfield  {title} {\enquote {\bibinfo {title} {{Photon-axion
  mixing and ultra-high-energy cosmic rays from BL Lac type objects - Shining
  light through the Universe}},}\ }\href {\doibase 10.1103/PhysRevD.84.125019}
  {\bibfield  {journal} {\bibinfo  {journal} {Phys. Rev.}\ }\textbf {\bibinfo
  {volume} {D84}},\ \bibinfo {pages} {125019} (\bibinfo {year} {2011})},\
  \Eprint {http://arxiv.org/abs/0901.4085} {arXiv:0901.4085 [astro-ph.HE]}
  \BibitemShut {NoStop}%
\bibitem [{\citenamefont {Hochmuth}\ and\ \citenamefont
  {Sigl}(2007)}]{Hochmuth:2007hk}%
  \BibitemOpen
  \bibfield  {author} {\bibinfo {author} {\bibfnamefont {Kathrin~A.}\
  \bibnamefont {Hochmuth}}\ and\ \bibinfo {author} {\bibfnamefont {Guenter}\
  \bibnamefont {Sigl}},\ }\bibfield  {title} {\enquote {\bibinfo {title}
  {{Effects of Axion-Photon Mixing on Gamma-Ray Spectra from Magnetized
  Astrophysical Sources}},}\ }\href {\doibase 10.1103/PhysRevD.76.123011}
  {\bibfield  {journal} {\bibinfo  {journal} {Phys. Rev.}\ }\textbf {\bibinfo
  {volume} {D76}},\ \bibinfo {pages} {123011} (\bibinfo {year} {2007})},\
  \Eprint {http://arxiv.org/abs/0708.1144} {arXiv:0708.1144 [astro-ph]}
  \BibitemShut {NoStop}%
\bibitem [{\citenamefont {Brax}\ and\ \citenamefont
  {Burrage}(2011)}]{Brax:2010gp}%
  \BibitemOpen
  \bibfield  {author} {\bibinfo {author} {\bibfnamefont {Philippe}\
  \bibnamefont {Brax}}\ and\ \bibinfo {author} {\bibfnamefont {Clare}\
  \bibnamefont {Burrage}},\ }\bibfield  {title} {\enquote {\bibinfo {title}
  {{Atomic Precision Tests and Light Scalar Couplings}},}\ }\href {\doibase
  10.1103/PhysRevD.83.035020} {\bibfield  {journal} {\bibinfo  {journal} {Phys.
  Rev.}\ }\textbf {\bibinfo {volume} {D83}},\ \bibinfo {pages} {035020}
  (\bibinfo {year} {2011})},\ \Eprint {http://arxiv.org/abs/1010.5108}
  {arXiv:1010.5108 [hep-ph]} \BibitemShut {NoStop}%
\bibitem [{\citenamefont {Brax}\ and\ \citenamefont
  {Zioutas}(2010)}]{Brax:2010xq}%
  \BibitemOpen
  \bibfield  {author} {\bibinfo {author} {\bibfnamefont {Philippe}\
  \bibnamefont {Brax}}\ and\ \bibinfo {author} {\bibfnamefont {Konstantin}\
  \bibnamefont {Zioutas}},\ }\bibfield  {title} {\enquote {\bibinfo {title}
  {{Solar Chameleons}},}\ }\href {\doibase 10.1103/PhysRevD.82.043007}
  {\bibfield  {journal} {\bibinfo  {journal} {Phys. Rev.}\ }\textbf {\bibinfo
  {volume} {D82}},\ \bibinfo {pages} {043007} (\bibinfo {year} {2010})},\
  \Eprint {http://arxiv.org/abs/1004.1846} {arXiv:1004.1846 [astro-ph.SR]}
  \BibitemShut {NoStop}%
\bibitem [{\citenamefont {Payez}\ \emph {et~al.}(2015)\citenamefont {Payez},
  \citenamefont {Evoli}, \citenamefont {Fischer}, \citenamefont {Giannotti},
  \citenamefont {Mirizzi},\ and\ \citenamefont {Ringwald}}]{Payez:2014xsa}%
  \BibitemOpen
  \bibfield  {author} {\bibinfo {author} {\bibfnamefont {Alexandre}\
  \bibnamefont {Payez}}, \bibinfo {author} {\bibfnamefont {Carmelo}\
  \bibnamefont {Evoli}}, \bibinfo {author} {\bibfnamefont {Tobias}\
  \bibnamefont {Fischer}}, \bibinfo {author} {\bibfnamefont {Maurizio}\
  \bibnamefont {Giannotti}}, \bibinfo {author} {\bibfnamefont {Alessandro}\
  \bibnamefont {Mirizzi}}, \ and\ \bibinfo {author} {\bibfnamefont {Andreas}\
  \bibnamefont {Ringwald}},\ }\bibfield  {title} {\enquote {\bibinfo {title}
  {{Revisiting the SN1987A gamma-ray limit on ultralight axion-like
  particles}},}\ }\href {\doibase 10.1088/1475-7516/2015/02/006} {\bibfield
  {journal} {\bibinfo  {journal} {JCAP}\ }\textbf {\bibinfo {volume} {1502}},\
  \bibinfo {pages} {006} (\bibinfo {year} {2015})},\ \Eprint
  {http://arxiv.org/abs/1410.3747} {arXiv:1410.3747 [astro-ph.HE]} \BibitemShut
  {NoStop}%
\bibitem [{\citenamefont {Burrage}\ \emph
  {et~al.}(2009{\natexlab{b}})\citenamefont {Burrage}, \citenamefont {Davis},\
  and\ \citenamefont {Shaw}}]{Burrage:2008ii}%
  \BibitemOpen
  \bibfield  {author} {\bibinfo {author} {\bibfnamefont {Clare}\ \bibnamefont
  {Burrage}}, \bibinfo {author} {\bibfnamefont {Anne-Christine}\ \bibnamefont
  {Davis}}, \ and\ \bibinfo {author} {\bibfnamefont {Douglas~J.}\ \bibnamefont
  {Shaw}},\ }\bibfield  {title} {\enquote {\bibinfo {title} {{Detecting
  Chameleons: The Astronomical Polarization Produced by Chameleon-like Scalar
  Fields}},}\ }\href {\doibase 10.1103/PhysRevD.79.044028} {\bibfield
  {journal} {\bibinfo  {journal} {Phys. Rev.}\ }\textbf {\bibinfo {volume}
  {D79}},\ \bibinfo {pages} {044028} (\bibinfo {year} {2009}{\natexlab{b}})},\
  \Eprint {http://arxiv.org/abs/0809.1763} {arXiv:0809.1763 [astro-ph]}
  \BibitemShut {NoStop}%
\bibitem [{\citenamefont {Olive}\ and\ \citenamefont
  {Pospelov}(2008)}]{Olive:2007aj}%
  \BibitemOpen
  \bibfield  {author} {\bibinfo {author} {\bibfnamefont {Keith~A.}\
  \bibnamefont {Olive}}\ and\ \bibinfo {author} {\bibfnamefont {Maxim}\
  \bibnamefont {Pospelov}},\ }\bibfield  {title} {\enquote {\bibinfo {title}
  {{Environmental dependence of masses and coupling constants}},}\ }\href
  {\doibase 10.1103/PhysRevD.77.043524} {\bibfield  {journal} {\bibinfo
  {journal} {Phys. Rev.}\ }\textbf {\bibinfo {volume} {D77}},\ \bibinfo {pages}
  {043524} (\bibinfo {year} {2008})},\ \Eprint {http://arxiv.org/abs/0709.3825}
  {arXiv:0709.3825 [hep-ph]} \BibitemShut {NoStop}%
\bibitem [{\citenamefont {Menezes}\ \emph {et~al.}(2005)\citenamefont
  {Menezes}, \citenamefont {Avelino},\ and\ \citenamefont
  {Santos}}]{Menezes:2005tp}%
  \BibitemOpen
  \bibfield  {author} {\bibinfo {author} {\bibfnamefont {J.}~\bibnamefont
  {Menezes}}, \bibinfo {author} {\bibfnamefont {P.~P.}\ \bibnamefont
  {Avelino}}, \ and\ \bibinfo {author} {\bibfnamefont {C.}~\bibnamefont
  {Santos}},\ }\bibfield  {title} {\enquote {\bibinfo {title} {{Varying alpha
  monopoles}},}\ }\href {\doibase 10.1103/PhysRevD.72.103504} {\bibfield
  {journal} {\bibinfo  {journal} {Phys. Rev.}\ }\textbf {\bibinfo {volume}
  {D72}},\ \bibinfo {pages} {103504} (\bibinfo {year} {2005})},\ \Eprint
  {http://arxiv.org/abs/hep-ph/0509326} {arXiv:hep-ph/0509326 [hep-ph]}
  \BibitemShut {NoStop}%
\bibitem [{\citenamefont {Baker}\ \emph {et~al.}(2013)\citenamefont {Baker}
  \emph {et~al.}}]{Baker:2013zta}%
  \BibitemOpen
  \bibfield  {author} {\bibinfo {author} {\bibfnamefont {K.}~\bibnamefont
  {Baker}} \emph {et~al.},\ }\bibfield  {title} {\enquote {\bibinfo {title}
  {{The quest for axions and other new light particles}},}\ }\href {\doibase
  10.1002/andp.201300727} {\bibfield  {journal} {\bibinfo  {journal} {Annalen
  Phys.}\ }\textbf {\bibinfo {volume} {525}},\ \bibinfo {pages} {A93--A99}
  (\bibinfo {year} {2013})},\ \Eprint {http://arxiv.org/abs/1306.2841}
  {arXiv:1306.2841 [hep-ph]} \BibitemShut {NoStop}%
\bibitem [{\citenamefont {Ringwald}(2014)}]{Ringwald:2014vqa}%
  \BibitemOpen
  \bibfield  {author} {\bibinfo {author} {\bibfnamefont {A.}~\bibnamefont
  {Ringwald}},\ }\bibfield  {title} {\enquote {\bibinfo {title} {{Axions and
  Axion-Like Particles}},}\ }in\ \href
  {https://inspirehep.net/record/1304441/files/arXiv:1407.0546.pdf} {\emph
  {\bibinfo {booktitle} {{Proceedings, 49th Rencontres de Moriond on
  Electroweak Interactions and Unified Theories}}}}\ (\bibinfo {year} {2014})\
  pp.\ \bibinfo {pages} {223--230},\ \Eprint {http://arxiv.org/abs/1407.0546}
  {arXiv:1407.0546 [hep-ph]} \BibitemShut {NoStop}%
\bibitem [{\citenamefont {Hamilton}\ \emph {et~al.}(2015)\citenamefont
  {Hamilton}, \citenamefont {Jaffe}, \citenamefont {Haslinger}, \citenamefont
  {Simmons}, \citenamefont {Müller},\ and\ \citenamefont
  {Khoury}}]{Hamilton:2015zga}%
  \BibitemOpen
  \bibfield  {author} {\bibinfo {author} {\bibfnamefont {Paul}\ \bibnamefont
  {Hamilton}}, \bibinfo {author} {\bibfnamefont {Matt}\ \bibnamefont {Jaffe}},
  \bibinfo {author} {\bibfnamefont {Philipp}\ \bibnamefont {Haslinger}},
  \bibinfo {author} {\bibfnamefont {Quinn}\ \bibnamefont {Simmons}}, \bibinfo
  {author} {\bibfnamefont {Holger}\ \bibnamefont {Müller}}, \ and\ \bibinfo
  {author} {\bibfnamefont {Justin}\ \bibnamefont {Khoury}},\ }\bibfield
  {title} {\enquote {\bibinfo {title} {{Atom-interferometry constraints on dark
  energy}},}\ }\href {\doibase 10.1126/science.aaa8883} {\bibfield  {journal}
  {\bibinfo  {journal} {Science}\ }\textbf {\bibinfo {volume} {349}},\ \bibinfo
  {pages} {849--851} (\bibinfo {year} {2015})},\ \Eprint
  {http://arxiv.org/abs/1502.03888} {arXiv:1502.03888 [physics.atom-ph]}
  \BibitemShut {NoStop}%
\end{thebibliography}%

\end{document}